\documentclass[prd,aps,letterpaper,twocolumn,preprintnumbers,nofootinbib,bibnotes,superscriptaddress,floatfix]{revtex4}

\usepackage[utf8]{inputenc}
\usepackage{amsmath}
\usepackage{amssymb}
\usepackage{mathtools}
\usepackage{slashed}
\usepackage{color}
\usepackage{comment}
\usepackage{subfigure}
\usepackage{graphicx}
\usepackage{dcolumn}
\usepackage{bm}
\usepackage[hypertexnames=false]{hyperref}
\usepackage[normalem]{ulem}
\newcommand{\nn}{\nonumber}
\DeclareRobustCommand{\Eq}[1]{Eq.~\eqref{eq:#1}}

\newcommand{\df}{\mathrm{d}}
\newcommand{\img}{\mathrm{i}}
\newcommand{\eps}{\epsilon}
\newcommand{\cA}{\mathcal{A}}
\newcommand{\cO}{\mathcal{O}}

\newcommand{\bt}{\v b_T}

\newcommand{\MS}{\overline{\mathrm{MS}}}
\newcommand{\RI}{\mathrm{RI}^\prime\mathrm{/MOM}}

\renewcommand{\v}[1]{\ensuremath{\mathbf{#1}}} 

\begin{document}

\preprint{MIT/CTP-5153}

\title{Nonperturbative renormalization of staple-shaped \\ Wilson line operators in lattice QCD}

\author{Phiala Shanahan}
 \email{phiala@mit.edu}
 \affiliation{Center for Theoretical Physics, Massachusetts Institute of Technology, Cambridge, MA 02139, USA}%
\author{Michael L. Wagman}%
 \email{mwagman@gmail.com}
 \affiliation{Center for Theoretical Physics, Massachusetts Institute of Technology, Cambridge, MA 02139, USA}%
\author{Yong Zhao}%
 \email{yzhao@bnl.gov}
 \affiliation{Center for Theoretical Physics, Massachusetts Institute of Technology, Cambridge, MA 02139, USA}%
 \affiliation{Physics Department, Brookhaven National Laboratory, Bldg. 510A, Upton, NY 11973, USA}%

\date{\today}
\begin{abstract}
Quark bilinear operators with staple-shaped Wilson lines are used to study transverse-momentum-dependent parton distribution functions (TMDPDFs) from lattice quantum chromodynamics (QCD). 
Here, the renormalization factors for the isovector operators, including all mixings between operators with different Dirac structures, are computed nonperturbatively in the regularization-independent momentum subtraction scheme for the first time. This study is undertaken in quenched QCD with three different lattice spacings. With Wilson flow applied to the gauge fields in the calculations, the operator mixing pattern due to chiral symmetry breaking with the lattice regularization is found to be significantly different from that predicted by one-loop lattice perturbation theory calculations. These results constitute a critical step towards the systematic extraction of TMDPDFs from lattice QCD.
\end{abstract}

\maketitle

\section{Introduction}

Building a quantitative description of the structure of the proton in terms of its fundamental parton constituents is a defining goal of hadronic physics research. A key aspect of this structure is encoded in transverse-momentum-dependent parton distribution functions (TMDPDFs), which describe the intrinsic transverse momentum of partons in the proton~\cite{Collins:1981uk,Collins:1981va,Collins:1984kg}. 
When the transverse momentum of the parton, $q_T$, is in the perturbative region of quantum chromodynamics (QCD), i.e. $q_T\gg \Lambda_{\rm QCD}$, the TMDPDFs can be obtained perturbatively in terms of collinear parton distribution functions (PDFs)~\cite{Collins:1981uw,Collins:1984kg}. 
In contrast, when $q_T\sim \Lambda_{\rm QCD}$, the TMDPDFs are intrinsically nonperturbative, and constraining these fundamental aspects of proton structure remains a challenging problem for both theory and experiment.

TMDPDFs can be determined experimentally via measurements of Drell-Yan production or semi-inclusive deep inelastic scattering (SIDIS) of electrons off nucleons; continued efforts aim to extract the distributions by fits to global experimental data~\cite{Landry:1999an,Landry:2002ix,Konychev:2005iy,Su:2014wpa,DAlesio:2014mrz,Echevarria:2014xaa,Kang:2015msa,Bacchetta:2017gcc,Scimemi:2017etj,Bertone:2019nxa}. Improved constraints on these quantities are expected from measurements at COMPASS~\cite{Gautheron:2010wva}, the 12~GeV program at the Thomas Jefferson National Accelerator Facility~\cite{Dudek:2012vr}, RHIC~\cite{Aschenauer:2015eha}, and a future Electron-Ion Collider~\cite{Accardi:2012qut}. Complementing the experimental efforts, recent progress also enables first-principles lattice QCD calculations of aspects of TMD physics. In particular, ratios of the Bjorken-$x$ moments of different TMDPDFs have been computed~\cite{Musch:2010ka,Musch:2011er,Engelhardt:2015xja,Yoon:2016dyh,Yoon:2017qzo}, and extensions of the large-momentum effective theory (LaMET) approach~\cite{Ji:2013dva,Ji:2014gla}, which was originally proposed to enable the $x$-dependence of PDFs to be constrained from lattice QCD, have been under study for the case of TMDPDFs~\cite{Ji:2014hxa,Ji:2018hvs,Ebert:2018gzl,Ebert:2019okf,Ebert:2019tvc,Ji:2019sxk}. 
Moreover, a systematic procedure to extract the Collins-Soper evolution kernel, which governs the energy evolution of TMDPDFs, has been established based on calculations of ratios of TMDPDFs from lattice QCD~\cite{Ebert:2018gzl,Ebert:2019okf,Ebert:2019tvc}.

Lattice QCD studies of TMDPDFs involve the nonperturbative  calculation of hadron matrix  elements of nonlocal bilinear operators with staple-shaped Wilson lines. These matrix elements are referred to as unsubtracted quasi TMDPDFs~\cite{Ji:2014hxa,Ji:2018hvs} or quasi beam functions~\cite{Ebert:2018gzl,Ebert:2019okf}.
An important component of such calculations is the renormalization of the bare quasi beam functions and their matching to the modified minimal subtraction ($\MS$) scheme.
Bare quasi beam functions display both logarithmic and linear ultraviolet (UV) divergences. The linear divergences originate from the self-energies of the Wilson lines, and can be absorbed into exponential  factors~\cite{Dorn:1986dt,Polyakov:1979gp,Dotsenko:1979wb,Craigie:1980qs,Brandt:1981kf,Knauss:1984rx,Korchemsky:1987wg,Korchemsky:1991zp}.
For hadronic matrix elements of quark bilinear operators with straight Wilson lines, which define the quasi PDFs, renormalization has been extensively studied in both perturbative and nonperturbative schemes~\cite{Dorn:1986dt,Craigie:1980qs,Stefanis:1983ke,Ji:2015jwa,Ishikawa:2016znu,Chen:2016fxx,Constantinou:2017sej,Alexandrou:2017huk,Chen:2017mzz,Green:2017xeu,Stewart:2017tvs,Izubuchi:2018srq,Liu:2018uuj,Spanoudes:2018zya} and multiplicative renormalizability in coordinate space has been proven to all orders in continuum perturbation theory~\cite{Ji:2017oey,Ishikawa:2017faj,Green:2017xeu,Zhang:2018diq,Li:2018tpe}. Similarly, it is expected that quark bilinear operators with staple-shaped Wilson lines are also multiplicatively renormalizable~\cite{Yoon:2017qzo,Constantinou:2019vyb,Ebert:2019tvc}, such that they can be renormalized nonperturbatively via the regularization-independent momentum subtraction scheme ($\RI$). The matching from $\RI$ to $\MS$ can then be calculated analytically in the continuum theory with dimensional regularization, which is free from linear divergences. The one-loop matching coefficient has been calculated for operators with zero longitudinal separation of the quark fields~\cite{Constantinou:2019vyb}, which are relevant in the study of the $x$-moments of the TMDPDFs~\cite{Musch:2010ka,Musch:2011er,Engelhardt:2015xja,Yoon:2016dyh,Yoon:2017qzo}, and also for operators with quark fields separated longitudinally, which determine the $x$-dependence of the TMDPDFs~\cite{Ebert:2019tvc}.

Here, the $\RI$ renormalization of quasi beam functions is studied numerically in quenched QCD with improved Wilson valence fermions. Due to the explicit breaking of Lorentz and chiral symmetries in the calculation, the multiplicatively-renormalizable quark bilinear operators with straight or staple-shaped Wilson lines mix with others with different Dirac structures~\cite{Constantinou:2017sej,Green:2017xeu,Chen:2017mie,Constantinou:2019vyb}; the complete 16$\times$16 operator mixing matrix for staple-shaped operators with all possible Dirac structures is therefore computed here for the first time. This study is undertaken at lattice spacings of 0.04, 0.06, and 0.08~fm, and with a single lattice volume, $L\sim 2$~fm. This enables a first analysis of the lattice-spacing dependence of the mixing patterns induced in the lattice theory. 

To increase the signal-to-noise ratio of the calculation, the renormalization factors are computed using gauge field configurations for which the gauge links have been subject to 100 steps of Wilson flow to flow-time $\mathfrak{t} = 1.0$~\cite{Luscher:2010iy}.
With the flow-time fixed in lattice units, Wilson flow corresponds to a smearing prescription whose effects vanish in the continuum limit~\cite{Bonati:2014tqa}.
A subset of the calculations were also repeated without flow applied to the gauge fields. 
Typically, this smearing prescription significantly reduces mixing between different operator structures.
This modifies the mixing patterns such that the dominant mixings are not necessarily those predicted by one-loop lattice perturbation theory with the unflowed action.
Calculations are predominantly undertaken with a light quark mass corresponding to a pion mass of $m_\pi\sim 1.2$~GeV.
On the coarsest lattice, calculations with $m_\pi\sim 340$~MeV are also undertaken. 
While naively one might expect the heavy quark mass to enhance operator mixing due to the chiral-symmetry breaking of the mass terms in the fermion action, little mass-dependence is observed in the results.

Finally, the subset of the nonperturbative $\RI$ renormalization factors required to calculate renormalized quark-bilinear operators with Dirac structure $\gamma_4$ are matched to the $\MS$ scheme using matching coefficients computed in one-loop perturbation theory~\cite{Ebert:2019tvc}, and their lattice-spacing dependence is studied. This completes the nonperturbative renormalization prescription for quasi beam functions needed to study TMDPDFs from lattice QCD.
The key result is that, for the action considered here, and with Wilson flow applied to the gauge fields in the calculations, the operator mixing pattern due to chiral symmetry breaking with the lattice regularization is found to be significantly different from that predicted by one-loop lattice perturbation theory. Complete calculations of renormalized quasi TMDPDFs in this framework will thus require the computation of a larger set of operator structures than one might naively expect from perturbative studies.

\section{Quasi tmdpdfs}

TMDPDFs, which are relevant to scattering processes such as Drell-Yan and SIDIS, can  be  expressed  in  terms  of  beam  functions (also referred to as unsubtracted TMDPDFs) which describe the incoming collinear partons in the scattering process, and soft functions, which encode the effects of soft gluon radiation by partons. Beam functions are defined as hadron matrix elements of quark bilinear operators with staple-shaped Wilson lines extended along the light-cone direction, while the soft functions are defined as the vacuum matrix elements of Wilson loops extended along the incoming and outgoing light-cone directions.
Since they are defined on the light-cone, neither the beam nor soft functions can be directly calculated with lattice QCD formulated in Euclidean space. 
Constraints on TMDPDFs from lattice QCD, however, are possible via the LaMET approach~\cite{Ji:2013dva,Ji:2014gla}.
The principle of LaMET is to approximate light-cone parton distributions by static quasi distributions, defined in terms of Euclidean matrix elements which can be calculated nonperturbatively in highly boosted hadron states. At large hadron momentum, quasi distributions are then matched to light-cone parton distributions perturbatively. To calculate TMDPDFs, quasi TMDPDFs have been constructed in terms of quasi beam and quasi soft functions~\cite{Ji:2014hxa,Ji:2018hvs,Ebert:2018gzl,Ebert:2019okf}. 
Due to the complication of the quasi soft function,\footnote{It was recently proposed in Ref.~\cite{Ji:2019sxk} that the soft function can be calculated from heavy quark effective theory or a light-meson form factor combined with two quasi TMD distribution amplitudes in lattice QCD.} the  relation  between  quasi TMDPDFs and TMDPDFs is expected to be nonperturbative; the explicit form of this relation was presented in Refs.~\cite{Ebert:2018gzl,Ebert:2019okf}. It was also shown in those works that contributions from the soft sector, which do not depend on the hadron state, cancel in certain ratios of TMDPDFs and in the corresponding ratios of quasi TMDPDFs. For this reason, physical observables defined by ratios of TMDPDFs can be determined from lattice QCD calculations of quasi beam functions alone.
For example, the Collins-Soper kernel can be obtained from ratios of quasi beam functions at different hadron momenta~\cite{Ebert:2018gzl,Ebert:2019okf}.
Similarly, ratios of the $x$-moments of TMDPDFs can also be determined with lattice QCD~\cite{Musch:2010ka,Musch:2011er,Engelhardt:2015xja,Yoon:2016dyh,Yoon:2017qzo}.

\begin{figure}
    \centering
    \includegraphics[width=0.7\columnwidth]{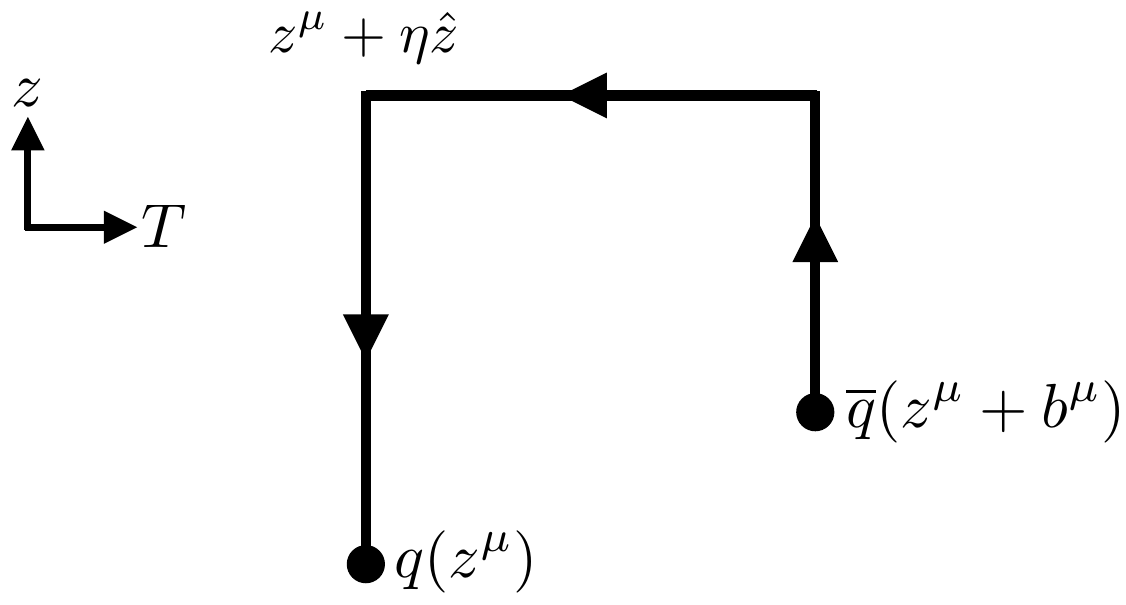}
    \caption{Illustration of the staple-shaped Wilson line structure of the quark bilinear operators defining quasi beam functions, as defined in Eq.~\eqref{eq:op}.}
    \label{fig:staple}
\end{figure}

Precisely, quasi beam functions are calculated as matrix elements of quark bilinear operators with staple-shaped Wilson lines, in position space: 
\begin{align} \label{eq:qbeam}
\tilde B^\Gamma_{q}(b^z, \bt,\eta,P^z)
=
\Bigl\langle h(P) \big| &\mathcal{O}_\Gamma^{q}(b^\mu,0,\eta) \big| h(P) \Bigr\rangle.
\end{align}
Here, $h(P)$ denotes a boosted hadron state with four-momentum $P^\mu$. The staple-shaped Wilson-line operator $\mathcal{O}_\Gamma^{q}(b^\mu,0,\eta)$ is built as a bilinear of quark flavor $q$, with a Wilson line of staple length $\eta$ in the $\hat{z}$ direction connecting endpoints separated by $b^\mu = b_z + b_T$, where $T$ denotes a direction transverse to $\hat{z}$: 
\begin{align}\nonumber
\mathcal{O}^{q}_\Gamma(b^\mu,z^\mu,\eta)\equiv &\bar q(z^\mu + b^\mu) \frac{\Gamma}{2} W_{\hat z}(z^\mu + b^\mu ;\eta-b^z) \\\nonumber
&\times 
W^\dagger_{T}(z^\mu + \eta \hat{z}; b_T) W^\dagger_{\hat z}(z^\mu;\eta) q(z^\mu)\\\label{eq:op}
\equiv&\bar q(z^\mu + b^\mu) \frac{\Gamma}{2}\widetilde{W}(\eta;b^\mu;z^\mu)q(z^\mu).
\end{align}
This operator, which is depicted graphically in Figure~\ref{fig:staple}, is constructed from spatial Wilson lines that are defined as 
\begin{align} \label{eq:coll_Wilson_L}
W_{\hat \alpha}(x^\mu;\eta) &= P \exp\left[ \img g \int_0^{\eta} \df s \, \cA^\alpha(x^\mu + s \hat \alpha) \right].
\end{align}

Fourier transforms with respect to $b^z$ of the quasi beam function for different Dirac structures $\Gamma \in$\{$\mathbb{I}$,$\gamma_\mu$,$\gamma_5$,$\gamma_\mu\gamma_5$,$\sigma_{\mu\nu}$\} define quasi TMDPDFs with different spin structures.
Including the quasi soft factor $\tilde{\Delta}_{S}^{q}$~\cite{Ji:2014hxa,Ji:2018hvs,Ebert:2018gzl,Ebert:2019okf}, the quasi TMDPDF in the $\overline{\rm MS}$ scheme is defined as
\begin{align}\label{eq:quasiTMD}
\tilde{f}_{q,\Gamma}^{\mathrm{TMD}}\big(x, \vec{b}_{T}, \mu, P^{z}\big)\equiv & \lim_{\eta\rightarrow\infty}
\int \frac{\mathrm{d} b^{z}}{2 \pi}\ e^{-\mathrm{i} b^{z}\left(x P^{z}\right)} \mathcal{Z}^{\overline{\text{MS}}}_{\Gamma\Gamma'}(\mu,b^z)\nonumber
\\&\times \frac{2P^z}{N_\Gamma} \!\tilde{B}^{\Gamma'}_{q}\big(b^{z},\! \vec{b}_{T},\! \eta,\! P^{z}\big) \tilde{\Delta}_{S}^{q}\left(b_{T},\! \eta\right)\,,
\end{align}
where $\mathcal{Z}^{\MS}_{\Gamma\Gamma'}(\mu,b^z)$ renormalizes the quasi TMDPDF and matches it to the $\MS$ scheme at scale $\mu$ and $N_\Gamma$ is a normalization factor needed to ensure covariance under the Euclidean analogs of Lorentz transformations. 
For the case $\Gamma = \gamma^4$ considered below, $N_{\gamma^4} = 1/(2 E_{\vec{P}})$ where $E_{\vec{P}}$ is the energy of the hadron state boosted to three-momentum $\vec{P}$. 
The details of the definition and other properties of the quasi soft factor $\tilde{\Delta}_{S}^{q}$ are omitted here; since it only depends on $b_T$, it will cancel in ratios of quasi TMDPDFs, as formed in key applications including calculations to extract the Collins-Soper kernel~\cite{Musch:2010ka,Musch:2011er,Engelhardt:2015xja,Yoon:2016dyh,Yoon:2017qzo,Ebert:2018gzl,Ebert:2019okf}.

Both the quasi beam function and soft factor have linear UV divergences that are proportional to the total length of the Wilson line in the associated operator, and as such the quasi soft factor also acts as a counterterm to cancel the linear divergences in the bare quasi beam function.
Nevertheless, there is still a remaining linear divergence proportional to $|b^z|$, as well as other logarithmic divergences, which are renormalized by $\mathcal{Z}_{\Gamma\Gamma'}^{\overline{\text{MS}}}$. {The renormalization factor $\mathcal{Z}_{\Gamma\Gamma'}^{\overline{\text{MS}}}$ can be separated into pieces which renormalize the quasi beam function and soft factor, $Z_{\cO_{\Gamma\Gamma'}}^{\overline{\text{MS}}}$ and $Z_{S}^{\overline{\text{MS}}}$, respectively:
\begin{equation}\label{eq:separatedZ}
\mathcal{Z}_{\Gamma\Gamma'}^{\overline{\text{MS}}}(\mu,b^z) = Z_{\cO_{\Gamma\Gamma'}}^{\overline{\text{MS}}}(\mu,b^z,\vec{b}_T,\eta) Z_{S}^{\overline{\text{MS}}}(\mu,b_T,\eta)\,.
\end{equation}
Since $Z_{S}^{\overline{\text{MS}}}(\mu,b_T,\eta)$ is independent of $b^z$, it will also be canceled in ratios of quasi TMDPDFs at $\eta$ and $b_T$, thus leaving only the quasi beam functions to be renormalized for key applications.
}
Taking into account mixing among $\mathcal{O}^{q}_\Gamma(b^\mu,z^\mu,\eta)$ with different Dirac structures, $Z^{\overline{\text{MS}}}_{\cO_{\Gamma\Gamma'}}$ is a 16$\times$16 matrix that can be computed nonperturbatively via the $\RI$ prescription~\cite{Martinelli:1994ty,Martinelli:1993dq} with a perturbative matching to the $\MS$ scheme~\cite{Ebert:2019tvc}, as detailed in the next section.

\section{nonperturbative renormalization}

The bare staple-shaped Wilson line operator, Eq.~\eqref{eq:op}, and hence the bare quasi beam function, can be renormalized via the nonperturbative $\RI$ prescription~\cite{Martinelli:1994ty,Martinelli:1993dq}.
In this approach, a renormalization constant is defined to relate the bare and tree-level amputated Green's functions for a given operator in a gauge-fixed quark or gluon state at a fixed scale. A perturbative matching calculated in continuum perturbation theory then relates the resulting $\RI$ renormalized operator to the $\overline{\text{MS}}$ scheme.
For a lattice operator $\mathcal{O}^\text{latt}_\Gamma$, which implicitly depends on the staple extent $\eta$, the displacement between the staple endpoints $b^\mu$, and the lattice spacing $a$, this renormalization can be expressed as a matrix equation accounting for mixing of operators with different Dirac structures $\Gamma$: 
\begin{align}\label{eq:renormfull}
   \mathcal{O}_\Gamma^{\overline{\text{MS}}}(\mu) =&\lim_{a\to0} \mathcal{R}_{\mathcal{O}_{\Gamma\Gamma''}}^{\overline{\text{MS}}}(\mu,p_R)Z^\text{$\RI$}_{\mathcal{O}_{\Gamma''\Gamma'}}(p_R,a)\mathcal{O}_{\Gamma'}^\text{latt}(a) \nonumber\\
   =&\lim_{a\to0} Z^{\overline{\text{MS}}}_{\mathcal{O}_{\Gamma\Gamma'}}(\mu,a)\mathcal{O}^\text{latt}_{\Gamma'}(a)\,, 
\end{align}
where $p_R$ denotes the matching scale introduced in the $\RI$ scheme.
The determination of the nonperturbative $\RI$ renormalization matrix $Z^\text{$\RI$}_{\mathcal{O}_{\Gamma\Gamma'}}(p_R,a)$ is discussed in Sec.~\ref{sec:RIMOM}, while the continuum perturbation theory calculation of the conversion factor $\mathcal{R}_{\mathcal{O}_{\Gamma\Gamma'}}^{\overline{\text{MS}}}(\mu,p_R)$ from the $\RI$ scheme to $\overline{\text{MS}}$ is outlined in Sec.~\ref{sec:matching}.

At all orders in perturbation theory, the scheme conversion factor $\mathcal{R}_{\mathcal{O}_{\Gamma\Gamma'}}^{\overline{\text{MS}}}(\mu,p_R)$ cancels the $p_R$ and gauge dependence of $Z^\text{$\RI$}_{\mathcal{O}_{\Gamma\Gamma'}}(p_R,a)$, and therefore renormalized matrix elements of $\mathcal{O}_\Gamma^{\overline{\text{MS}}}$ only depend on $\eta$, $b^\mu$, and the $\MS$ scale $\mu$. Typically, however, $\mathcal{R}_{\mathcal{O}_{\Gamma\Gamma'}}^{\overline{\text{MS}}}(\mu,p_R)$ is calculated at finite orders in perturbation theory; for the operators considered here, only one-loop results have been computed~\cite{Ebert:2019tvc}, so the cancellation is incomplete. 
Moreover, at finite $a$ there are lattice artifacts that have $p_R$ dependence.
In the pseudoscalar case ($\Gamma=\gamma_5$), $Z^\text{RI/MOM}_{\mathcal{O}_{\gamma_5\Gamma'}}(p_R,a)$ additionally develops a nonperturbative Goldstone boson pole that depends on $p_R$~\cite{Cudell:1998ic}. 

For the operators considered here, the one-loop corrections in the diagonal terms $\mathcal{R}_{\mathcal{O}_{\Gamma\Gamma}}^{\overline{\text{MS}}}(\mu,p_R)$ (i.e., with $\Gamma=\Gamma'$) are significantly larger than one~\cite{Ebert:2019tvc}, indicating that the perturbative series does not converge well. These large one-loop corrections can be canceled by combining $\mathcal{R}_{\mathcal{O}_{\Gamma\Gamma'}}^{\overline{\text{MS}}}(\mu,p_R)$ with the quasi soft factor $\tilde{\Delta}_S^q$ in the $\MS$ scheme, thus rendering the matching coefficient close to one~\cite{Ebert:2019tvc} without affecting ratios of quasi TMDPDFs or the extraction of key physics results determined by such ratios.
In the analysis presented here, higher-order perturbative corrections are neglected, and remnant $p_R$ dependence is treated as a discretization effect leading to systematic uncertainty in the results discussed in Sec.~\ref{sec:numerical}.
More details of discretization effects are discussed in Appendix~\ref{sec:disceffects}.

\subsection{$\RI$ scheme in lattice QCD}
\label{sec:RIMOM}

The matrix of $\RI$ renormalization constants $Z^\text{$\RI$}_{\mathcal{O}_{\Gamma\Gamma'}}$, for the quark bilinear operators with staple-shaped Wilson lines as defined in Eq.~\eqref{eq:op}, is defined by the condition 
\begin{equation}\label{eq:RIMOM}
   Z_q^{-1}(p_R) Z^\text{$\RI$}_{\mathcal{O}_{\Gamma\Gamma'}}(p_R)\Lambda^{\mathcal{O}_{\Gamma'}}_{\alpha\beta}(p) \big|_{p^\mu=p^\mu_R}=\Lambda_{\alpha\beta}^{\mathcal{O}_{\Gamma};\text{tree}}(p)\,,
\end{equation}
relating the bare and tree-level values of the operator's amputated Green's function in an off-shell quark state in the Landau gauge:
\begin{equation}
   \Lambda^{\mathcal{O}_{\Gamma}} (p) =  S^{-1} (p) G^{\mathcal{O}_{\Gamma}}(p) S^{-1} (p) \,,
\end{equation}
where $G^{\mathcal{O}_{\Gamma}}$ denotes the Green's function for operator $\mathcal{O}_\Gamma$ with Dirac structure $\Gamma$, which implicitly depends on the staple extent $\eta$ and displacement between staple endpoints $b^\mu$, and $S(p)$ is the quark propagator projected to momentum $p$. 
All quantities appearing on the left-hand-side of Eq.~\eqref{eq:RIMOM} implicitly depend on the lattice spacing; this dependence is suppressed in the following discussion.
In Eq.~\eqref{eq:RIMOM}, $\sqrt{p_R^2}$ acts a non-perturbative renormalization scale; however, since the operator $\mathcal{O}^{q}_{\Gamma}$ is nonlocal and frame dependent, the magnitude of $p_R^\mu$ alone is not sufficient to specify the renormalization condition. 
Different directions in $p_R^\mu$ amount to different renormalization schemes, which are related by finite renormalization factors. As a result, $Z^\text{$\RI$}_{\mathcal{O}_{\Gamma\Gamma'}}(p_R)$ depends on $p_R^\mu$ rather than only its magnitude.

In a calculation with lattice volume $V=L^3\times T$ and lattice spacing $a$, the non-amputated quark-quark Green's function with one insertion of the operator $\cO_\Gamma$ is 
\begin{equation} \label{Gdef}
   G^{\mathcal{O}_{\Gamma}}_{\alpha\beta} (p) =  \frac{1}{V}\sum_{x,y,z} {\rm e}^{ {\mathrm i} 
	p \cdot (x-y) } \langle q_\alpha (x) \cO_\Gamma (z+b,z) \bar{q}_\beta (y) \rangle,
\end{equation}
calculated as
\begin{equation} \label{Gdefexplicit}
   G^{\mathcal{O}_{\Gamma}}_{\alpha\beta} (p)\! =\! \frac{1}{V}\sum_{z} 
\langle\! \gamma_5 S^\dagger(p,\!b+z) \gamma_5 \widetilde{W}(\eta;b+z,\!z)\frac{\Gamma}{2} S(p,\!z) \!\rangle_{\alpha\beta},
\end{equation}
using the quark propagator
\begin{align}
S_{\alpha\beta}(p,x)&=\sum_y e^{-ip\cdot y}\langle  q_\alpha(x)\bar q_\beta(y)\rangle, \\
S_{\alpha\beta}(p) &= \frac{1}{V}\sum_x e^{ip\cdot x} S_{\alpha\beta}(p,x).
\end{align}
The quark wavefunction renormalization $Z_q$ is defined via
\begin{align}\label{eq:Zq}
&Z_q(p_R) S(p)\big|_{p^2=p_R^2} = S^\text{tree}(p) \\
\implies &Z_q(p_R) = \frac{1}{12}\text{Tr}\left[S^{-1}(p)S^\text{tree}(p)\right]\bigg|_{p^2=p_R^2},
\end{align}
computed as
\begin{equation} \label{defzq}
Z_q(p_R)  = \frac{ {\rm Tr} \left[ {\rm i} \sum_\lambda \gamma_\lambda 
	\sin (a p_\lambda)  S^{-1} (p) \right] }
{12 \sum_\lambda \sin^2 (a p_\lambda) } \bigg|_{p^2=p_R^2}.
\end{equation}
In terms of the projected vertex function 
\begin{equation}
   \mathcal{V}^{\mathcal{O}_{\Gamma\Gamma'}}(p)\equiv \text{Tr} \left[  \Lambda^{\mathcal{O}_{\Gamma}}(p) \Gamma' \right],
\end{equation}
the $\RI$ condition in Eq.~\eqref{eq:RIMOM}, for an operator $\mathcal{O}_\Gamma$ with endpoints separated by $b^\mu$, can be expressed as
\begin{align}\label{eq:RIMOM2}
   &Z_q^{-1}(p_R)Z^\text{$\RI$}_{\mathcal{O}_{\Gamma\Gamma''}}(p_R) \mathcal{V}^{\mathcal{O}_{\Gamma''\Gamma'}}(p)\big|_{p^\mu=p^p_R}\nn\\ &= \text{Tr}\left[ \Lambda^{\mathcal{O}_{\Gamma}}_\text{tree}(p_R) \Gamma' \right]= 6e^{i p_R\cdot b}\delta^{\Gamma\Gamma'},
\end{align}
which yields an expression for the matrix of renormalization factors at $p_R$:
\begin{equation}\label{eq:ZO}
   \left(Z^\text{$\RI$}_{\mathcal{O}_{\Gamma\Gamma'}}(p_R)\right)^{-1}  = \frac{\mathcal{V}^{\mathcal{O}_{\Gamma\Gamma'}}(p)}{6e^{i p_R\cdot b}Z_q(p_R) }\bigg|_{p^\mu=p^p_R} .
\end{equation}

\subsection{Conversion to the $\MS$ scheme}
\label{sec:matching}

Since the renormalized matrix element in the $\RI$ scheme is independent of the UV regulator, it differs from the result in the continuum limit only by discretization effects at finite lattice spacing. The $\RI$ matrix element can thus be matched to the $\MS$ scheme in continuum perturbation theory, and then extrapolated to the continuum limit using nonperturbative calculations at different values of $a$.

Elements of the matrix of matching coefficients $\mathcal{R}_{\mathcal{O}_{\Gamma \Gamma'}}^{\overline{\text{MS}}}(\mu,p_R)$ in Eq.~\eqref{eq:renormfull} have been calculated at one-loop order in continuum perturbation theory with dimensional regularization ($D=4-2\eps$) for operators $\mathcal{O}_\Gamma$ with both $b^z=0$~\cite{Constantinou:2019vyb} and $b^z\neq0$~\cite{Ebert:2019tvc}. This matching matrix can be expressed as
\begin{align}
   \mathcal{R}_{\mathcal{O}_{\Gamma \Gamma'}}^{\overline{\text{MS}}}(\mu,p_R)=Z_{\cO}^{\overline{\rm MS}} (\eps,\mu)\left[\tilde{Z}^{\rm \RI}_{\cO} (p_R,\mu,\eps)^{-1}\right]_{\Gamma\Gamma'}\,,
\end{align}
where $\tilde{Z}^{\rm \RI}_{\cO;\Gamma\Gamma''} (p_R,\mu,\eps)$ is the perturbatively-computed $\RI$ renormalization factor for the quasi beam function, defined in \Eq{ZO}. The scales $\mu$ and $p_R$ must be chosen to be large enough to satisfy $\mu, p_R \gg \Lambda_{QCD}$ in order to permit $\mathcal{R}_{\mathcal{O}_{\Gamma \Gamma'}}^{\overline{\text{MS}}}(\mu,p_R)$ to be accurately calculated in perturbation theory, while $p_R$ must simultaneously be taken to be much smaller than the lattice cutoff $\pi/a$ for discretization effects to be neglected. The factor $Z^{\overline{\rm MS}}(\epsilon,\mu)$ is gauge-invariant and universal for all Dirac structures $\Gamma$~\cite{Constantinou:2019vyb,Ebert:2019tvc}:
\begin{align}
   Z_{\cO}^{\overline{\rm MS}}(\epsilon,\mu) = 1- {\alpha_s c_f\over 4\pi}{7\over \epsilon} + O(\alpha_s^2)\,,
\end{align}
where $c_f=4/3$.

For $\Gamma=\gamma^\lambda$, the matching coefficient $\mathcal{R}_{\gamma^\lambda \Gamma'}^{\overline{\text{MS}}}$ has been calculated for all projectors $\Gamma'$ at one-loop order~\cite{Ebert:2019tvc}. The results are summarized here for completeness:
\begin{align}
    &\mathcal{R}_{\gamma^\lambda, \bf 1}^{\overline{\text{MS}}}(\mu,p_R) =  \mathcal{R}_{\gamma^\lambda, \gamma^5}^{\overline{\text{MS}}}(\mu,p_R)= \mathcal{R}_{\gamma^\lambda, \sigma^{\mu\nu}}^{\overline{\text{MS}}}(\mu,p_R)=0\,,\\
    &\mathcal{R}_{\gamma^\lambda, \gamma^\rho}^{\overline{\text{MS}}}(\mu,p_R) = 1\!+\! \left[{{\cal V}_{\gamma^\lambda,\gamma^\rho}^{(1)}(p_R,\mu)\over 6e^{i p_R\cdot b}}\!-\! Z_q^{(1)}(p_R,\mu)\delta_{\lambda \rho}\right]\,,\\
    &\mathcal{R}_{\gamma^\lambda, \gamma^\rho\gamma_5}^{\overline{\text{MS}}}(\mu,p_R) = {{\cal V}_{\gamma^\lambda,\gamma^\rho\gamma_5}^{(1)}(p_R,\mu)\over 6e^{i p_R\cdot b}}\,,
\end{align}
where $Z_q^{(1)}(p_R,\mu) = 0$
in the Landau gauge, and the subtraction of $1/\epsilon$ poles is implied. The explicit expression for the one-loop projected vertex functions ${\cal V}_{\gamma^\lambda,\gamma^\rho/\gamma^\rho\gamma_5}^{(1)}$ can be found in Ref.~\cite{Ebert:2019tvc}.

Defined in this way, the numerical values of the matching coefficients for the parameters of typical lattice QCD studies are much larger than one, which is due to $\eta/b_T$ terms that correspond to the rapidity divergences in the TMDPDF~\cite{Ji:2018hvs,Ebert:2019okf}. In \Eq{quasiTMD}, the quasi TMDPDF is defined with a quasi soft factor $\tilde{\Delta}_S^q = 1/\sqrt{S_q}$ which cancels the linear power divergences as well as the $\eta/b_T$ dependence in the quasi beam function; redefining the matching coefficient to include the quasi soft factor removes this divergence and yields a matching coefficient close to one\footnote{Note that unlike the proposal in Ref.~\cite{Ebert:2019tvc}, here the redefined matching coefficient $\tilde{\mathcal{R}}^{\overline{\text{MS}}}_{\Gamma \Gamma'}$ is not expanded as a series in $\alpha_s$. This maintains the numerical equivalence to $\MS$ matching when calculating ratios of quasi TMDPDFs.
   Moreover,  with Eq.~\eqref{eq:rprime} expanded in $\alpha_s$ as in Ref.~\cite{Ebert:2019tvc}, the quasi soft factor matching term only contributes to diagonal entries $\tilde{\mathcal{R}}^{\overline{\text{MS}}}_{\Gamma \Gamma}$, which leads to significantly enhanced operator mixing in $\MS$ results that is not present in $\RI$ results.
 }~\cite{Ebert:2019tvc}:
\begin{align}\label{eq:rprime}
    \tilde{\mathcal{R}}^{\overline{\text{MS}}}_{\Gamma \Gamma'}= & \frac{\mathcal{R}_{\Gamma \Gamma'}^{\overline{\text{MS}}}}{\sqrt{S_q}} = \frac{\mathcal{R}_{\Gamma \Gamma'}^{\overline{\text{MS}}}}{1+{\alpha_s c_f\over 4\pi} S_q^{(1)}}
    \,.
\end{align}
In the numerical investigation presented in the following section, the ``bent'' quasi soft factor defined in Ref.~\cite{Ebert:2019okf,Ebert:2019tvc} is adopted for this redefinition:
\begin{align}
    S^{\rm bent (1)}_q(b_T,\mu,\eta)\nn =&6\ln{\mu^2b_T^2\over 4e^{-2\gamma_E}}\!+\!12\!-\!4\ln{b_T^2+\eta^2\over \eta^2}\\
    &+8{\eta\over b_T}\arctan{\eta\over b_T}\nn+ {4b_T\over \sqrt{2}\eta}\arctan{b_T\over \sqrt{2}\eta} \\
    &- 2\ln{b_T^2+2\eta^2\over 2\eta^2}\,.
\end{align}
Since the quasi soft factor only depends on the operator staple geometry in terms of $b_T$ and $\eta$, its inclusion will not change the $p_R$ or $b^z$-dependence of the matching coefficient and therefore will not affect results for ratios of $\MS$ quasi beam functions at fixed $b_T$ and $\eta$.

\section{Numerical investigation} 
\label{sec:numerical}

The $\RI$ renormalization of quark bilinear operators with staple-shaped Wilson lines is studied on three quenched QCD ensembles, detailed in Table~\ref{tab:ensembles}. These ensembles are tuned to have lattice spacings of 0.04, 0.06, and 0.08~fm, and a common lattice volume, $L\sim 2$~fm. This enables study of the lattice-spacing dependence of renormalization factors and operator mixing patterns in the lattice theory. On each ensemble, $Z^{\RI}_{\mathcal{O}_{\Gamma\Gamma'}}(p)$ is computed via Eq.~\eqref{eq:ZO}, for the isovector combination of quark operators defined with staple extents $\eta$ ranging between 0.6--0.8~fm (specified in Table~\ref{tab:ensembles}), i.e., to almost half the lattice extent, and with staple widths and asymmetries $b_T$ and $b^z$ ranging from $-\eta$ to $\eta$, for the complete 16$\times$16 matrix of Dirac structures $\Gamma,\Gamma'$. The gauge link fields used in the calculation have been subjected to Wilson flow to flow-time $\mathfrak{t} = 1.0$~\cite{Luscher:2010iy}, to enhance the signal-to-noise ratio in the numerical results;{\footnote{In this calculation the flowed gauge fields were also used for constructing $\slashed{D}$.}} a study of the impact of this smearing prescription on mixing patterns is given in Appendix~\ref{app:flow}.  Valence quark propagators are computed with the tree-level $O(a)$ improved Wilson clover fermion action~\cite{Sheikholeslami:1985ij} with $\kappa$ values as given in Table~\ref{tab:ensembles}; these choices correspond to a pion mass of 1.2~GeV on each ensemble. For the $E_{24}$ ensemble, propagators corresponding to a pion mass of 340~MeV are also computed to enable a study of the mass-dependence of the renormalization patterns.
Ten different momenta of the quark state are considered, tabulated in Table~\ref{tab:moms}. While the dependence of the $\RI$ renormalization on the matching scales $p_R$ and $p_R^z$ would be canceled by an all-orders matching to the $\MS$ scheme, residual dependence on these scales remains with a matching calculated perturbatively to one-loop order. Studying various momenta at a range of scales $p_R^2$ from 
5.7 to 28~GeV$^2$ and $p_R^z$ from 1.3 to 2.6~GeV
allows an assessment of the systematic uncertainties in this matching.

\begin{table}[t]
	\begin{tabular}{ccccccc}\hline\hline
		Label & $\beta$ & $a$ [fm] & $L^3\times T$ & $\eta$ & $\kappa$ & $n_\text{cfg}$\\\hline
		$E_{24}$ & 6.1005 & 0.08 & $24^3\times 48$ & 7,9,11 & 0.121,0.1248 & 30 \\
		$E_{32}$ & 6.3017 & 0.06 & $32^3\times 64$ & 10,12,14 &0.1222 & 30 \\
		$E_{48}$ & 6.5977 & 0.04 & $48^3\times 96$ & 15,18,21 & 0.1233 & 10 \\ \hline\hline
	\end{tabular}
	\caption{\label{tab:ensembles}Ensembles of quenched QCD gauge field configurations used in this work~\cite{Detmold:2018zgk,Endres:2015yca}. $\beta$ values were chosen in Ref.~\cite{Asakawa:2015vta} to maintain a fixed physical volume, and $n_\text{cfg}$ configurations are analyzed on each ensemble.
      $L$, $T$, and $\eta$ are given in lattice units, where $\eta$ denotes the staple extents of the staple-shaped Wilson line operators (Eq.~\eqref{eq:op}) which are computed.  Valence quark propagators are computed with the tabulated $\kappa$ values, which correspond to pion masses consistent with $m_\pi = 1.20(5)$~GeV, on each ensemble, and for the $E_{24}$ ensemble additionally $m_\pi=340(20)~$MeV, on gauge fields subjected to Wilson flow as described in the text.
	}
\end{table}

\begin{table}[t]
    \centering
    \begin{tabular}{c|ccc}\hline\hline
       $n^\mu$  & $\sqrt{p^2}$ [GeV]  & $p^z$ [GeV] & $p^{[4]}/(p^2)^2$ \\ \hline
    (2,2,2,2) & 2.4 & 1.3 & 0.27 \\
    (2,2,2,4) & 2.7 & 1.3 & 0.25 \\
    (2,2,2,6) & 3.1 & 1.3 & 0.31 \\ \hline
    (3,3,3,2) & 3.5 & 1.9 & 0.30 \\
    (3,3,3,4) & 3.7 & 1.9 & 0.26 \\
    (3,3,3,6) & 4.0 & 1.9 & 0.25 \\
    (3,3,3,8) & 4.3 & 1.9 & 0.28 \\\hline
    (4,4,4,4) & 4.7 & 2.6 & 0.28 \\
    (4,4,4,6) & 4.9 & 2.6 & 0.26\\
    (4,4,4,8) & 5.2 & 2.6 & 0.25
   \\\hline\hline
    \end{tabular}
    \caption{Four-momenta considered in this work, where $p^\mu$ is the four-momentum in physical units corresponding to $n^\mu$ in lattice units. Results for $\RI$ renormalization factors are computed with $b_T < 0$ and are equivalent to results with $b_T > 0$ and a sign change in the component of $p_\mu$ along the axis used to define $b_T$. Final $\MS$ results only depend on $|b_T|$ up to neglected two-loop renormalization scheme matching effects; for convenience all results are given as functions of $|b_T|$ throughout.  Note that $p^\mu$ for a given $n^\mu$ is the same in physical units on all three ensembles. The H(4) invariant $p^{[4]} = \sum_{\mu=1}^4 p_\mu^4$ is discussed in Appendix~\ref{sec:disceffects}.}
    \label{tab:moms}
\end{table}

\subsection{Operator mixing with lattice regularization}
\label{subsec:mixingpatterns}

\begin{figure}
    \centering
    \includegraphics[width=0.9\columnwidth]{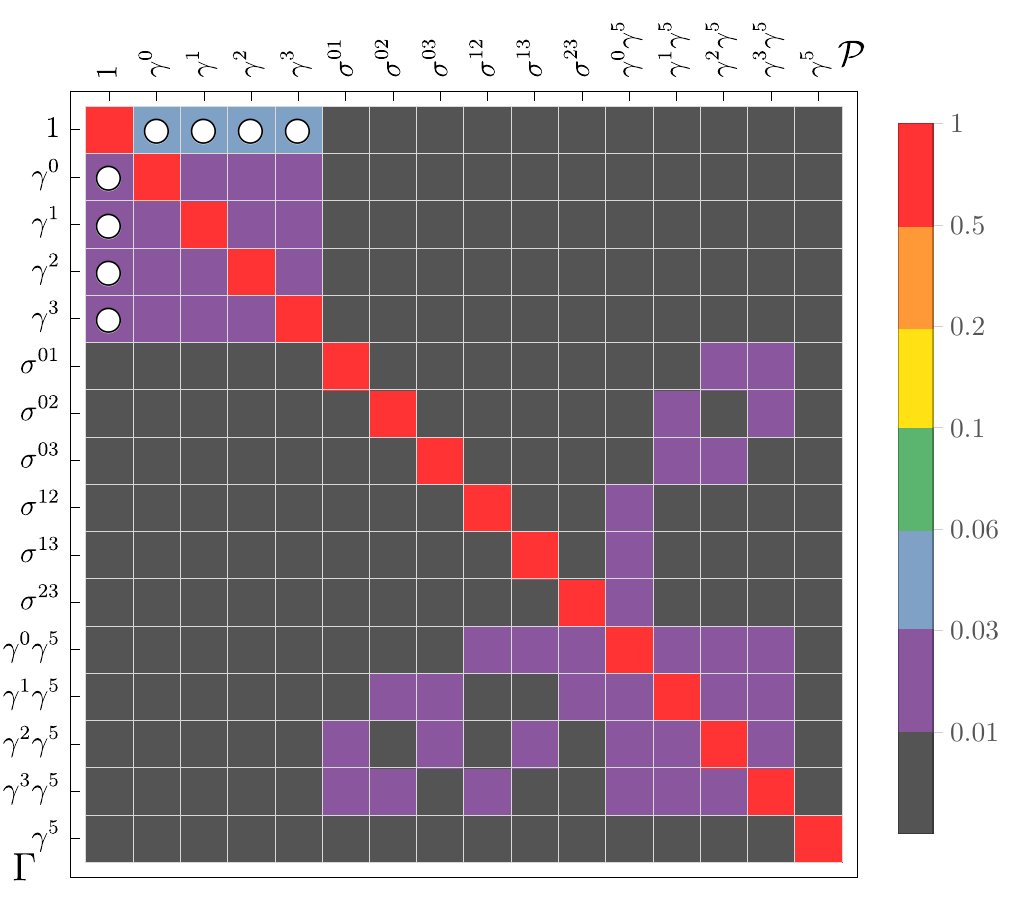}
    \caption{$\RI$ mixing pattern $\mathcal{M}^\text{$\RI$}_{\mathcal{O}_{\Gamma\mathcal{P}}}$ (Eq.~\eqref{eq:mixeq}) for local quark bilinear operators, calculated on the $E_{32}$ ensemble. White circles indicate the pattern of mixings predicted based on the off-shell nature of the quark in the relevant Green's functions~\cite{Martinelli:1994ty}.}
    \label{fig:local}
\end{figure}

\begin{figure}
    \centering
    \includegraphics[width=\columnwidth]{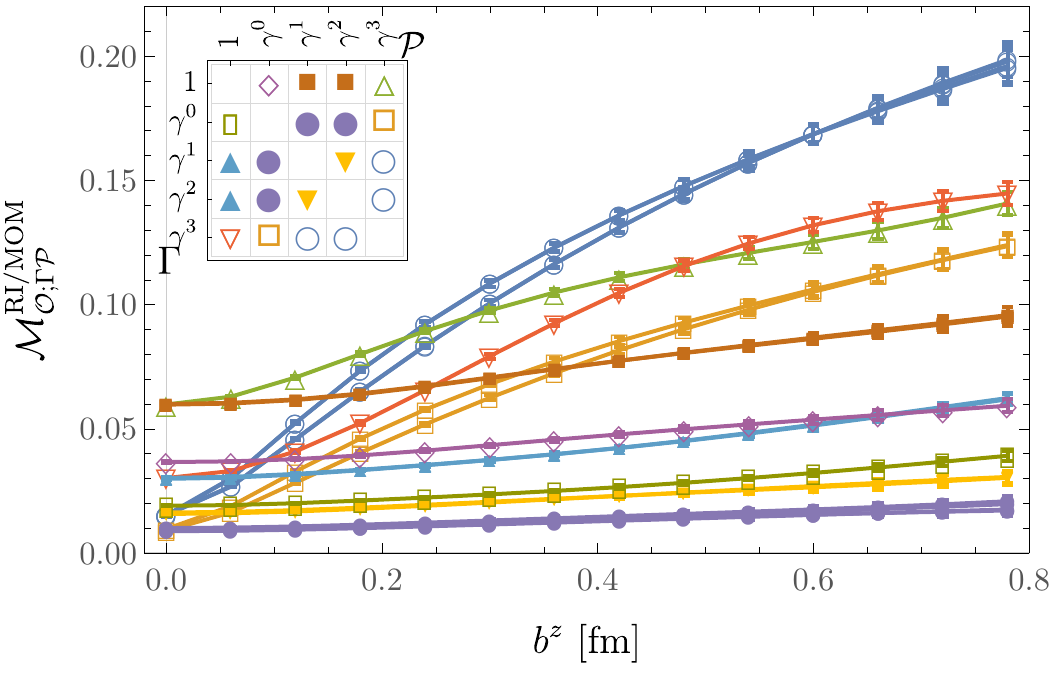}
    \caption{Submatrix of the $\RI$ mixing matrix $\mathcal{M}^\text{$\RI$}_{\mathcal{O}_{\Gamma\mathcal{P}}}(p_R)$ (Eq.~\eqref{eq:mixeq}) for quark bilinear operators with straight Wilson lines ($b_T=0$) with various extents $b^z$, for momentum $n^\nu=(2,2,2,2)$ in lattice units, calculated on the $E_{32}$ ensemble.}
    \label{fig:straightvsbz}
\end{figure}

\begin{figure*}
    \centering
    \includegraphics[width=\textwidth]{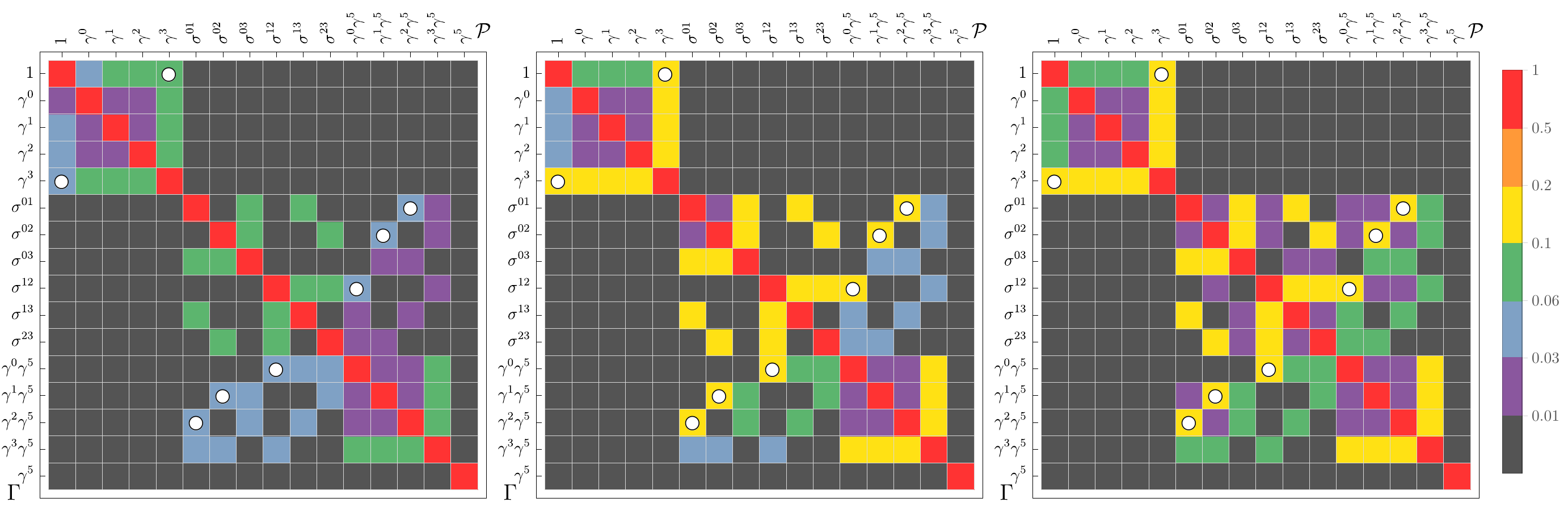}
    \caption{$\RI$ mixing pattern $\mathcal{M}^\text{$\RI$}_{\mathcal{O}_{\Gamma\mathcal{P}}}$ (Eq.~\eqref{eq:mixeq}) for quark bilinear operators with straight Wilson lines ($b_T=0$), calculated on the $E_{32}$ ensemble. The three panels, from left to right, show results for operator extents $b^z/a=\{3,7,11\}$. White circles indicate the mixings predicted by one-loop lattice perturbation theory and symmetry arguments~\cite{Constantinou:2017sej,Chen:2017mie,Green:2017xeu}.}
    \label{fig:straight}
\end{figure*}

\begin{figure*}
    \centering
    \includegraphics[width=\textwidth]{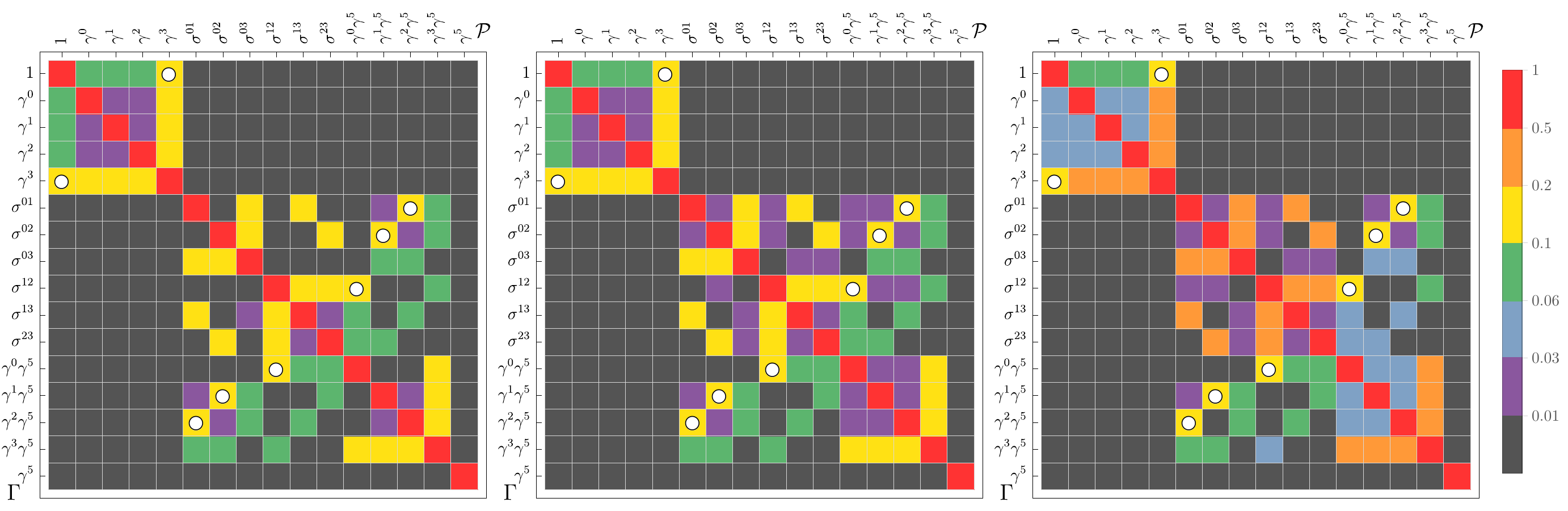}
    \caption{$\RI$ mixing pattern $\mathcal{M}^\text{$\RI$}_{\mathcal{O}_{\Gamma\mathcal{P}}}$ (Eq.~\eqref{eq:mixeq}) for quark bilinear operators with straight Wilson lines ($b_T=0$). From left to right, panels show results for operators with extent $b^z/a=11$ calculated on the ensembles $E_{24}$, $E_{32}$, $E_{48}$, with progressively finer lattice spacing $a$. White circles indicate the mixings predicted by one-loop lattice perturbation theory and symmetry arguments~\cite{Constantinou:2017sej,Chen:2017mie,Green:2017xeu}.}
    \label{fig:straightspacing}
\end{figure*}

\begin{figure*}
    \centering
    \includegraphics[width=\textwidth]{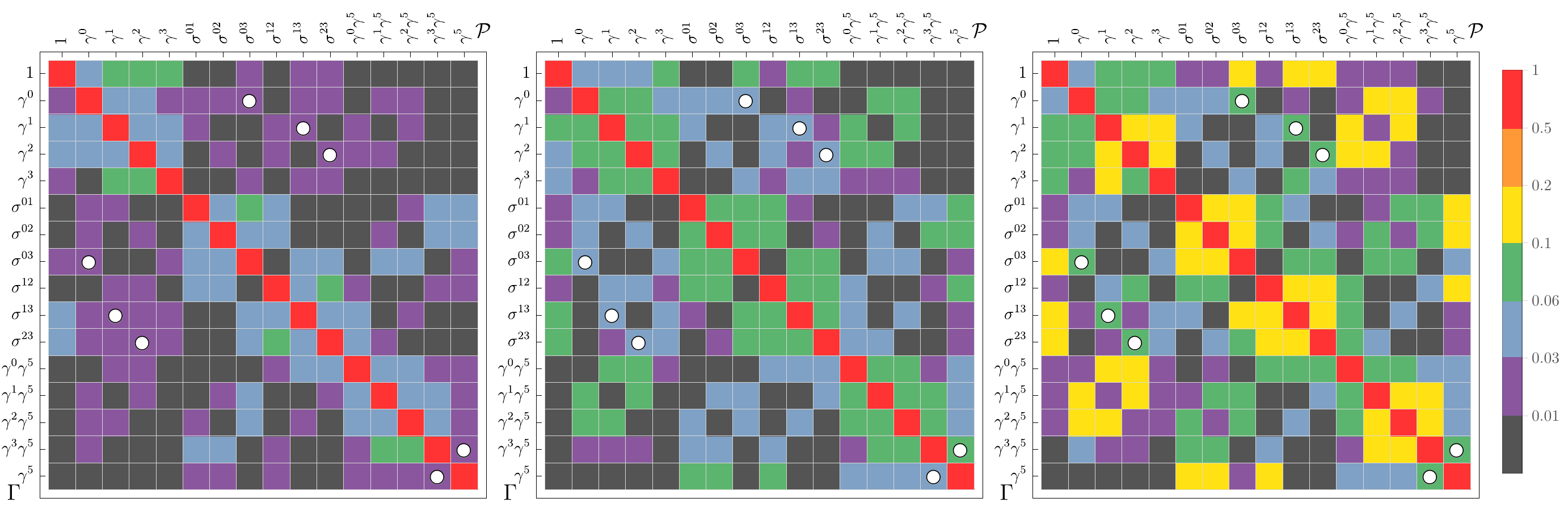}
    \caption{$\RI$ mixing pattern $\mathcal{M}^\text{$\RI$}_{\mathcal{O}_{\Gamma\mathcal{P}}}$ (Eq.~\eqref{eq:mixeq}) for quark bilinear operators with symmetric ($b^z=0$) staple-shaped Wilson lines. From left to right, panels show results for operators with $b_T/a=\{3,7,11\}$ and $\eta/a=14$, calculated on the $E_{32}$ ensemble. White circles indicate the mixings predicted by one-loop lattice perturbation theory~\cite{Constantinou:2019vyb}.}
    \label{fig:symstaple}
\end{figure*}

\begin{figure*}
    \centering
    \includegraphics[width=\textwidth]{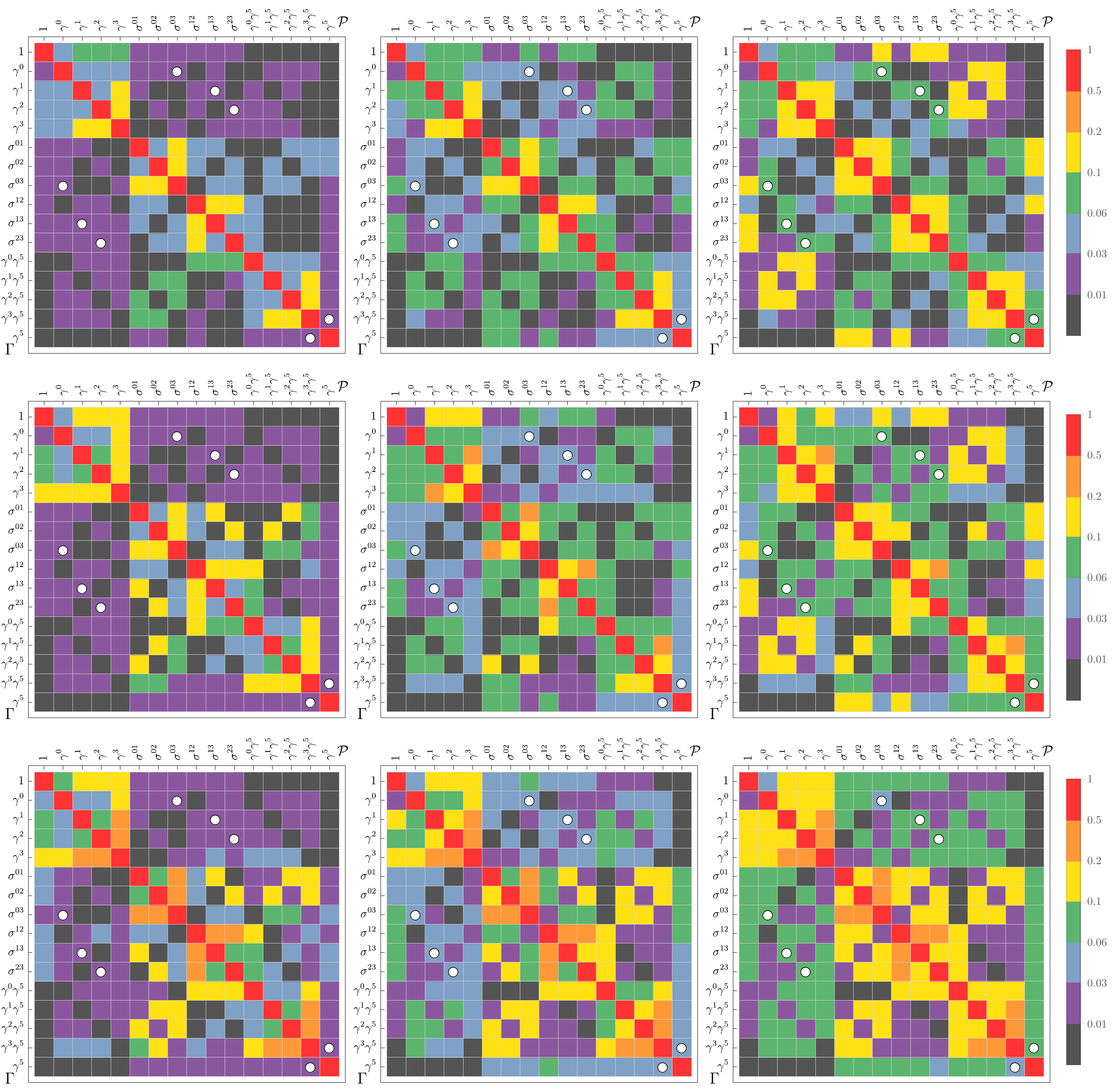}
    \caption{$\RI$ mixing pattern $\mathcal{M}^\text{$\RI$}_{\mathcal{O}_{\Gamma\mathcal{P}}}$ (Eq.~\eqref{eq:mixeq}) for quark bilinear operators with asymmetric staple-shaped Wilson lines ($b^z,b_T\neq0$) and $\eta/a=14$, calculated on the $E_{32}$ ensemble. From left to right, panels show results for operators with $b_T/a=\{3,7,11\}$, and from top to bottom with $b^z/a=\{3,7,11\}$. 
    For the asymmetric staple, there are no predictions available for the mixing patterns from one-loop lattice perturbation theory.}
    \label{fig:asymstaple1}
\end{figure*}

Ultimately, $\MS$ renormalization factors are computed by combining nonperturbatively-calculated $\RI$ factors with the one-loop perturbative matching to the $\MS$ scheme described in Sec.~\ref{sec:matching}. Comparison of the mixing patterns revealed in the matrix of nonperturbative $\RI$ factors with the patterns predicted by perturbation theory, which have been studied in the special cases of local operators, straight Wilson-line operators, and symmetric staple-shaped Wilson line operators, provides an indication of the important nonperturbative mixings for each operator.

Figures~\ref{fig:local}--\ref{fig:asymstaple1} display graphically the $16\times 16$ matrices of $\RI$ renormalization factors for all Dirac structures $\Gamma$ and projectors $\mathcal{P}$, for a range of operators with different staple widths and asymmetries $b_T$ and $b^z$, defined in Eq.~\eqref{eq:op}. In each case, percentage mixings relative to the average diagonal element are displayed, defined as:
\begin{align}\label{eq:mixeq}
   \mathcal{M}^\text{$\RI$}_{\mathcal{O}_{\Gamma\mathcal{P}}}=&\max_{p_R} \mathcal{M}^\text{$\RI$}_{\mathcal{O}_{\Gamma\mathcal{P}}}(p_R) \nn\\
   \equiv&\max_{p_R} \frac{\text{Abs}[Z^\text{$\RI$}_{\mathcal{O}_{\Gamma\mathcal{P}}}(p_R)]}{\frac{1}{16}\sum_i \text{Abs}[Z^\text{$\RI$}_{\mathcal{O}_{\Gamma_i\Gamma_i}}(p_R)]}\,,
\end{align}
where to illustrate the importance of mixings the maximum over momenta $p_R$ is taken over the ten momenta tabulated in Table~\ref{tab:moms}. Due to the off-shell nature of the quark in the Green's functions and the noncovariance of the operator $\mathcal{O}_\Gamma^{q}(b^\mu,0,\eta)$ itself, there can be  contributions from additional Dirac structures involving $p_R^\mu$ and $b^\mu$  to the vertex function of $\mathcal{O}_\Gamma^{q}(b^\mu,0,\eta)$, which do not break chiral symmetry and are also seen in continuum perturbation theory~\cite{Martinelli:1994ty}. In lattice calculations, there are also operator mixings arising from the breaking of chiral symmetry from the UV regularization that have been studied using one-loop lattice perturbation theory~\cite{Constantinou:2017sej,Constantinou:2019vyb}, auxiliary field methods~\cite{Green:2017xeu}, and symmetry arguments~\cite{Chen:2017mie}. 
On Figure~\ref{fig:local} and Figures~\ref{fig:straight}-\ref{fig:asymstaple1}, the predicted mixing patterns  are highlighted for comparison with the numerical results.\footnote{The operator mixing pattern for non-local quark bilinear operators with generic Wilson lines extending between them only depends on the directions of the Wilson lines at the endpoints~\cite{Green:2017xeu,Constantinou:2019vyb} and applies for asymmetric ($b^z \neq 0$) as well as symmetric staples. We thank Jeremy Green for discussions on this point.}

In general, operators with longer Wilson lines are seen to suffer from greater mixing effects than operators with shorter Wilson lines; this is shown explicitly for the straight Wilson line operators in Figure~\ref{fig:straightvsbz}.
Typically, the mixings predicted by lattice perturbation theory are found to be significant nonperturbatively, but in many cases other chiral-symmetry-preserving mixings are found to be equally, or more, important. For the matrix elements of quark bilinear operators with straight Wilson lines, the nonperturbative mixing pattern is seen to be block-diagonal for all Wilson line extents. Thus, while including only the operators expected to mix from perturbative arguments in a calculation of unsubtracted quasi PDFs would neglect important contributions, it is clear that not every Dirac structure must be considered.
The mixing patterns for matrix elements of quark bilinear operators with staple-shaped Wilson lines, however, are far more extensive and dense in the space of operator mixings, and almost every operator structure must be computed to renormalize a calculation of the unsubtracted quasi TMDPDFs in this framework. 

This suggests that additional operator mixings, for instance chiral-symmetry-breaking mixing between $\gamma^0$ and $\bf 1$, arise either from the interplay between symmetry-breaking operator mixing due to the lattice regularization and off-diagonal vertex function contributions involving $p_R^\mu$ and $b^\mu$, or from some additional mechanism. 
This mixing might be reduced by using lattice fermion actions with approximate chiral symmetry or by choosing a different definition of the vertex function used to define the $\RI$ operator renormalization condition, and results in this work on the relative importance of off-diagonal renormalization factors are specific to the $\RI$ scheme described in Sec.~\ref{sec:RIMOM}, which has been used in previous studies of nonlocal operators~\cite{Alexandrou:2017huk,Chen:2017mzz}.

The patterns of mixings computed on the three ensembles with different lattice spacings are consistent for each operator shape, with the relative magnitude of off-diagonal mixings relatively larger on the finer ensembles, as shown for the straight Wilson line case in Figure~\ref{fig:straightspacing}.
Studying the dependence of this mixing pattern on the choice of lattice action, including the effects of quenching, and nonperturbative renormalization scheme is left to future work. Future studies including a second lattice volume will also enable an investigation of the volume-dependence~\cite{Briceno:2018lfj} of the observed mixing patterns, although it is expected that finite-volume effects in the renormalization factors are smaller than finite-volume effects in hadron matrix elements involving the same operator~\cite{Alexandrou:2019lfo}.

A subset of calculations on the $E_{24}$ ensemble were repeated without Wilson flow applied to the gauge fields or Dirac operator.
As outlined in Appendix~\ref{app:flow}, Wilson flow generically reduces operator mixing, and in particular reduces some off-diagonal elements of the renormalization matrix significantly more than others, such that one-loop lattice perturbation theory (with an unflowed action) describes the unflowed mixing pattern somewhat better than the flowed mixing pattern.

\subsection{Renormalization results} \label{sec:numerics}

The row of the $\MS$ renormalization matrix for bare quasi beam functions with operator Dirac structure $\Gamma=\gamma_4$ is sufficient to determine $\MS$-renormalized matrix elements of $\mathcal{O}_{\gamma_4}^{\overline{\text{MS}}}$, given bare matrix elements $\mathcal{O}_\Gamma^{{\text{latt}}}$ for all 16 choices of $\Gamma$. 
These $\MS$ renormalization factors are defined from the nonperturbatively computed $\RI$ factor and the perturbative one-loop matching by
\begin{equation}\label{eq:MSrenorm}
   Z^{\overline{\text{MS}}}_{\mathcal{O}_{\gamma_4\Gamma}}(\mu,p_R) = \tilde{\mathcal{R}}_{\gamma_4\Gamma'}^{\overline{\text{MS}}}(\mu,p_R)Z^\text{$\RI$}_{\mathcal{O}_{\Gamma'\Gamma}}(p_R),
\end{equation}
where the left hand side is independent of the choice of $p_R$ up to  discretization effects, nonperturbative effects that vanish at asymptotically large $p_R^2$, and neglected two-loop perturbative matching corrections. 
Here, $Z^{\overline{\text{MS}}}_{\mathcal{O}_{\gamma_4\Gamma}}(\mu,p_R)$ implicitly includes the quasi soft factor included in $\tilde{\mathcal{R}}_{\gamma_4\Gamma'}^{\overline{\text{MS}}}(\mu,p_R)$ (and thus differs from $Z^{\overline{\text{MS}}}_{\mathcal{O}_{\gamma_4\Gamma}}(\mu,p_R)$ defined in Eq.~\eqref{eq:separatedZ} by terms which cancel in suitable ratios of renormalized TMDPDFs), and both $Z^{\overline{\text{MS}}}_{\mathcal{O}_{\gamma_4\Gamma}}(\mu,p_R)$ and $Z^\text{$\RI$}_{\mathcal{O}_{\Gamma'\Gamma}}(p_R)$ implicitly depend on $a$.
This renormalization factor is computed for each choice of $\Gamma$ with each of the 10 $p_R$ shown in Table~\ref{tab:moms}, for staple-shaped operators with $-\eta < b_T < \eta$, $-\eta \leq b^z \leq \eta$, for three values of $\eta$ on each ensemble shown in Table~\ref{tab:ensembles}. 

\begin{figure}[!t]
    \centering
    \includegraphics[width=\columnwidth]{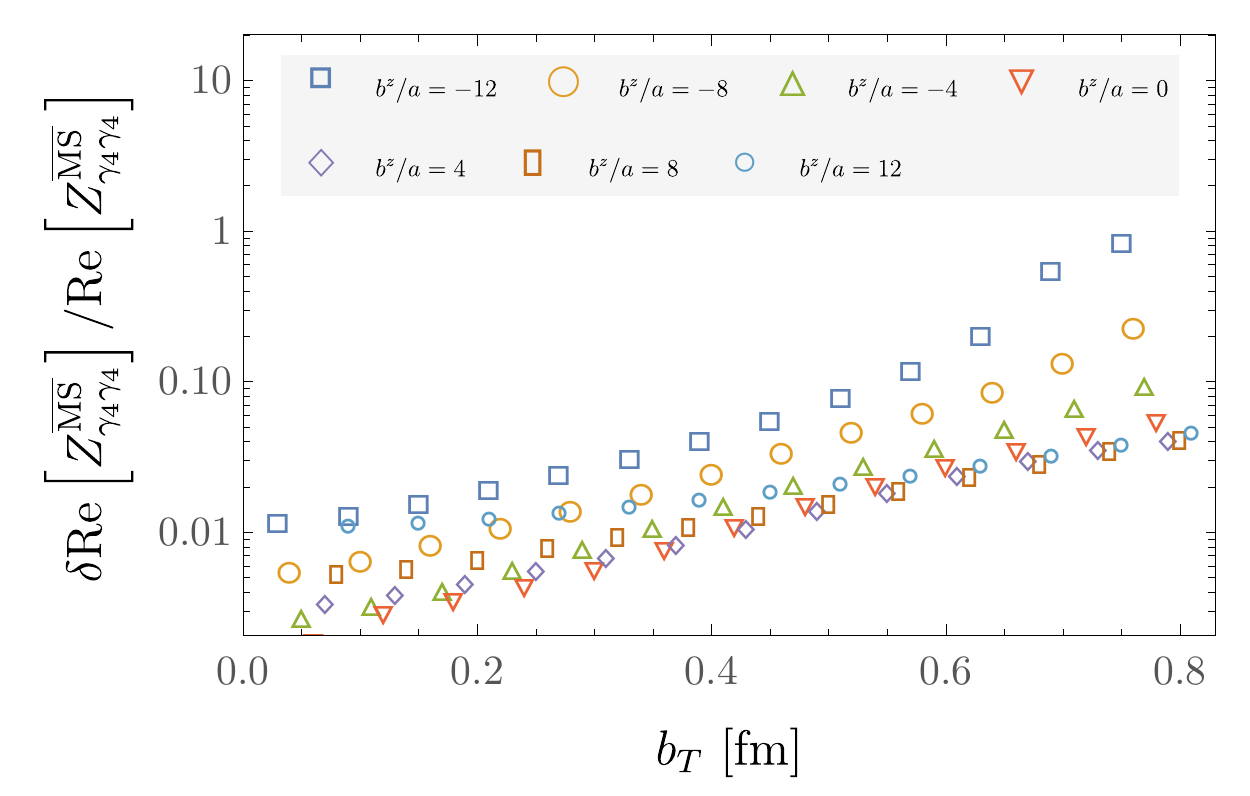}
    \caption{Scaling of the statistical noise in the nonlocal operator renormalization with the extent of the Wilson line. The noise-to-signal ratio shown increases approximately exponentially with the length of the Wilson line, and for example to achieve 5\% uncertainties on $Z^{\overline{\text{MS}}}_{\mathcal{O}_{\gamma_4\gamma_4}}(\mu = 2~\text{GeV})$ for a symmetric staple-shaped operator with $b_T = 1$ fm would require a statistical ensemble approximately ten times larger than the one used in this work.
    \label{fig:noise} }
\end{figure}

To determine $Z^{\overline{\text{MS}}}_{\mathcal{O}_{\gamma_4\Gamma}}$ from numerical results at different choices of $p_R$, one could fit the data to a model of the discretization effects in the renormalization matrix. However, statistical noise in the nonlocal operator renormalization grows exponentially with the length of the Wilson line  as illustrated in Figure~\ref{fig:noise}; in the present study it is not possible to constrain discretization effects from the 10 momenta used for all but the smallest nonlocal operator separations. 
In particular Bayes and Akaike information criteria prefer constant fits to more complicated fit forms including the leading discretization artifacts in the data (the functional form of these effects is made explicit in Appendix~\ref{sec:disceffects}). 
Moreover, the covariance matrices for nonlocal operators are not reliably estimated from the current data.

\begin{figure}[!t]
    \centering
    \includegraphics[width=\columnwidth]{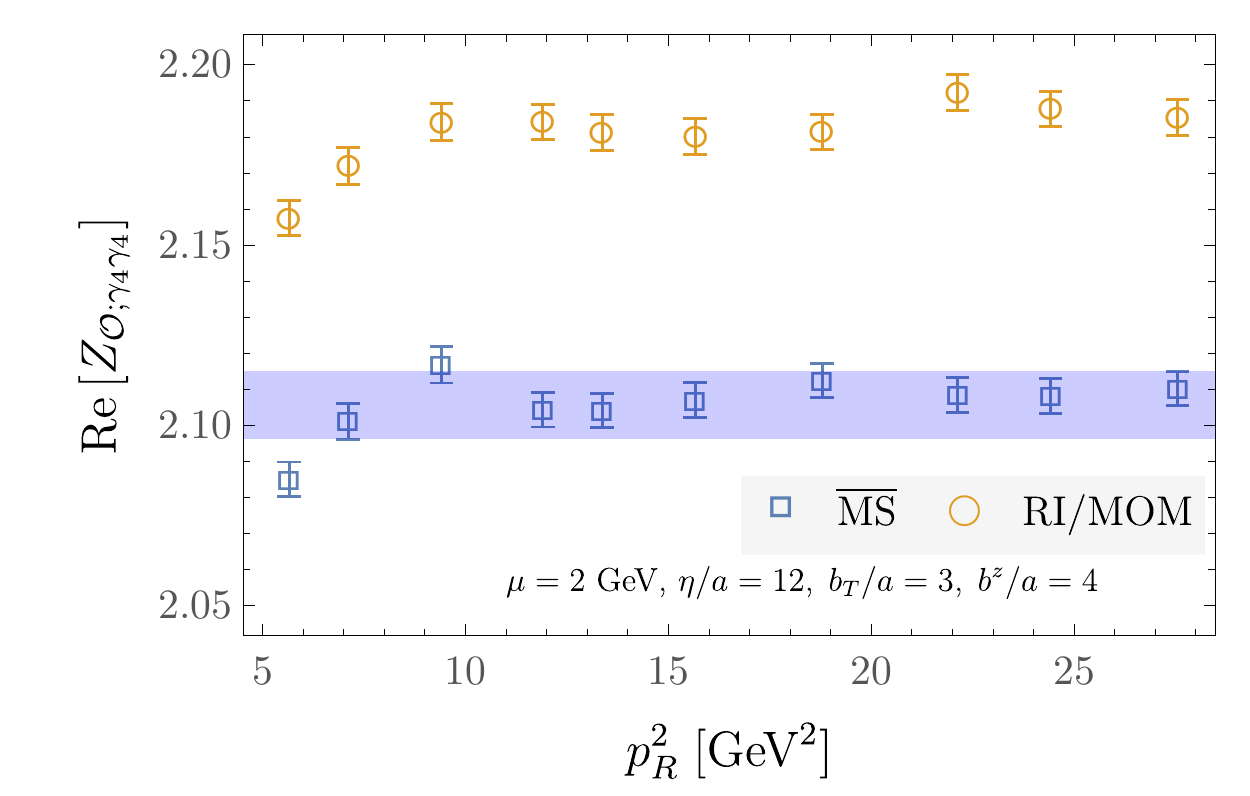}
    \caption{Numerical results for $Z^{\text{$\RI$}}_{\mathcal{O}_{\gamma_4\gamma_4}}(p_R)$ and $Z^{\MS}_{\mathcal{O}_{\gamma_4\gamma_4}}(\mu,p_R)$ for the $E_{32}$ ensemble with $\eta/a = 12$, $b^z/a = 4$, $b_T/a = 3$, $\mu = 2$ GeV, are displayed as orange circles and blue squares, respectively. Results at ten choices of $p_R$ given in Table~\ref{tab:moms} are shown. The blue shaded band shows the result of the weighted average in Eq.~\eqref{eq:weightedave} for $Z^{\MS}_{\mathcal{O}_{\gamma_4\gamma_4}}(\mu) \pm \delta Z^{\MS}_{\mathcal{O}_{\gamma_4\gamma_4}}(\mu)$.
    \label{fig:matching} }
\end{figure}

\begin{figure*}
    \subfigure[]{
        \centering
        \includegraphics[width=0.46\textwidth]{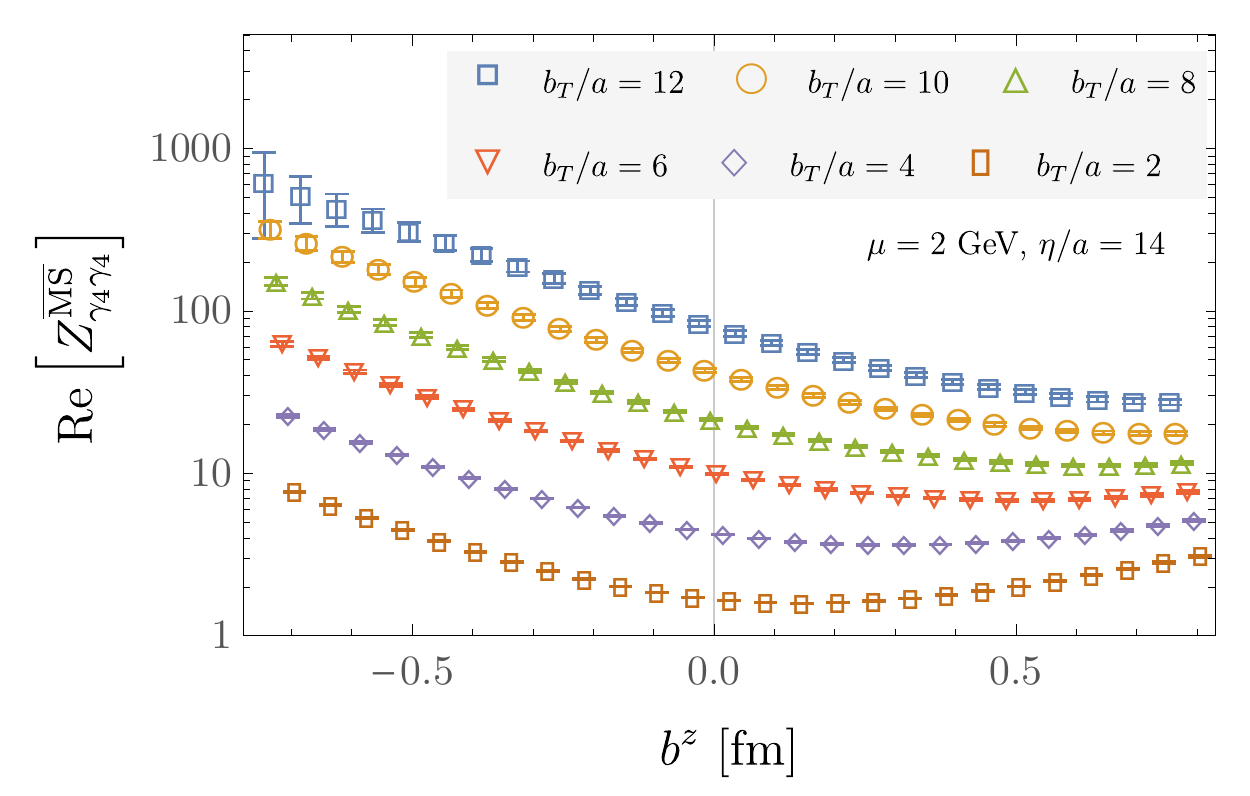}
        \label{fig:ZRvsBz}
        }\quad
    \subfigure[]{
        \centering
        \includegraphics[width=0.46\textwidth]{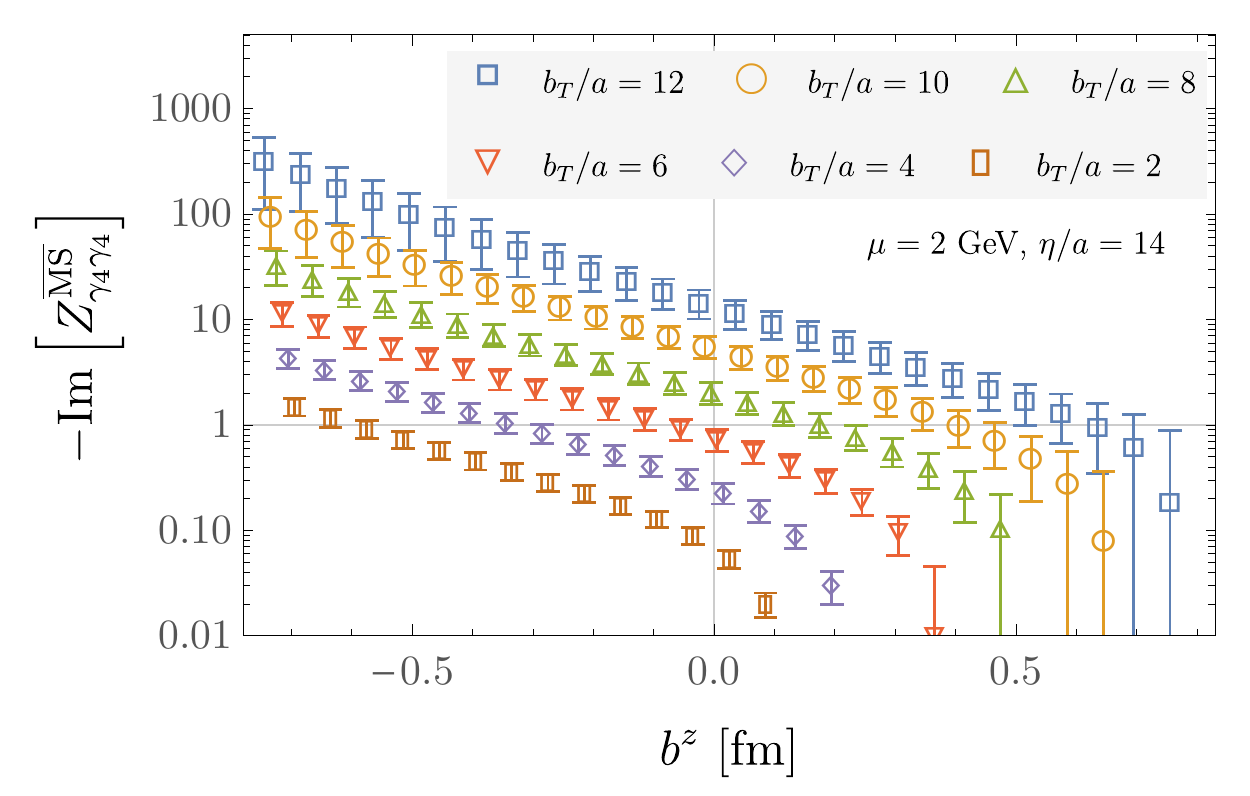}
        \label{fig:ZIvsBz} 
        }
    \subfigure[]{
        \centering
        \includegraphics[width=0.46\textwidth]{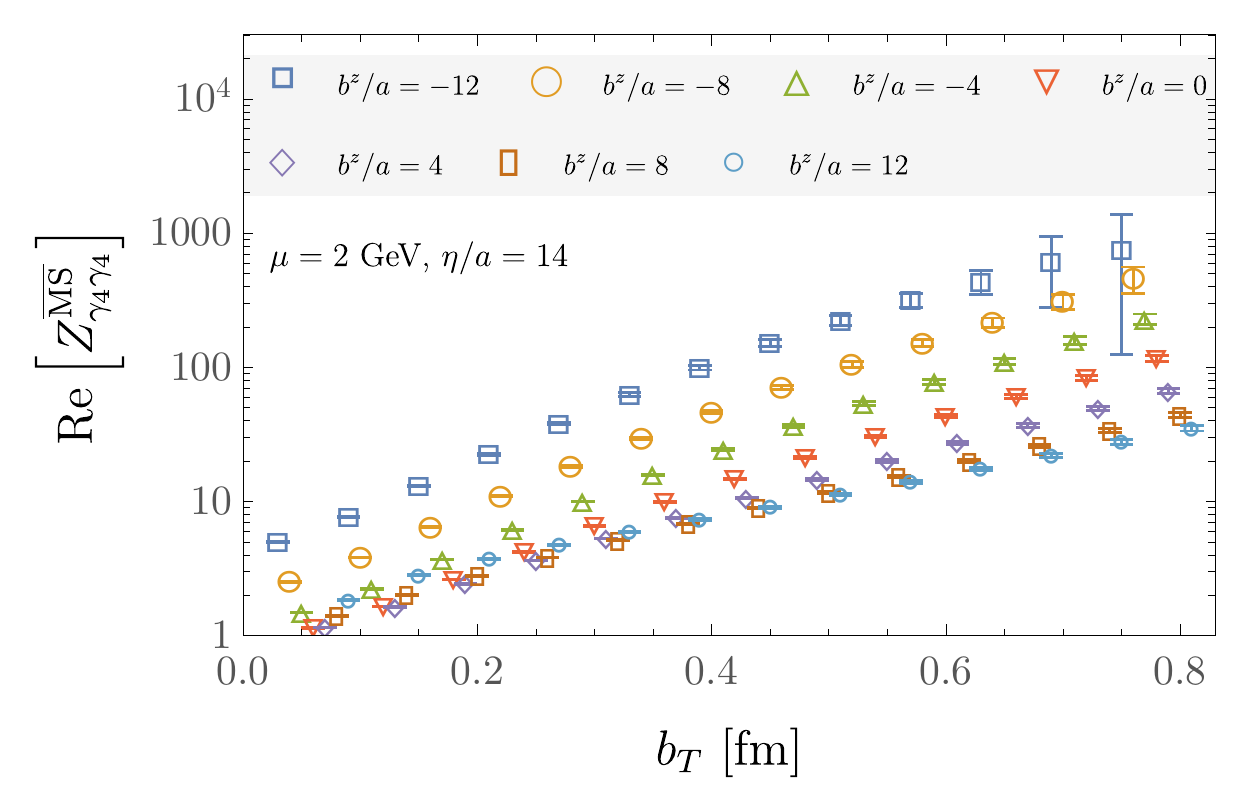}
        \label{fig:ZRvsBT}
        }\quad
    \subfigure[]{
        \centering
        \includegraphics[width=0.46\textwidth]{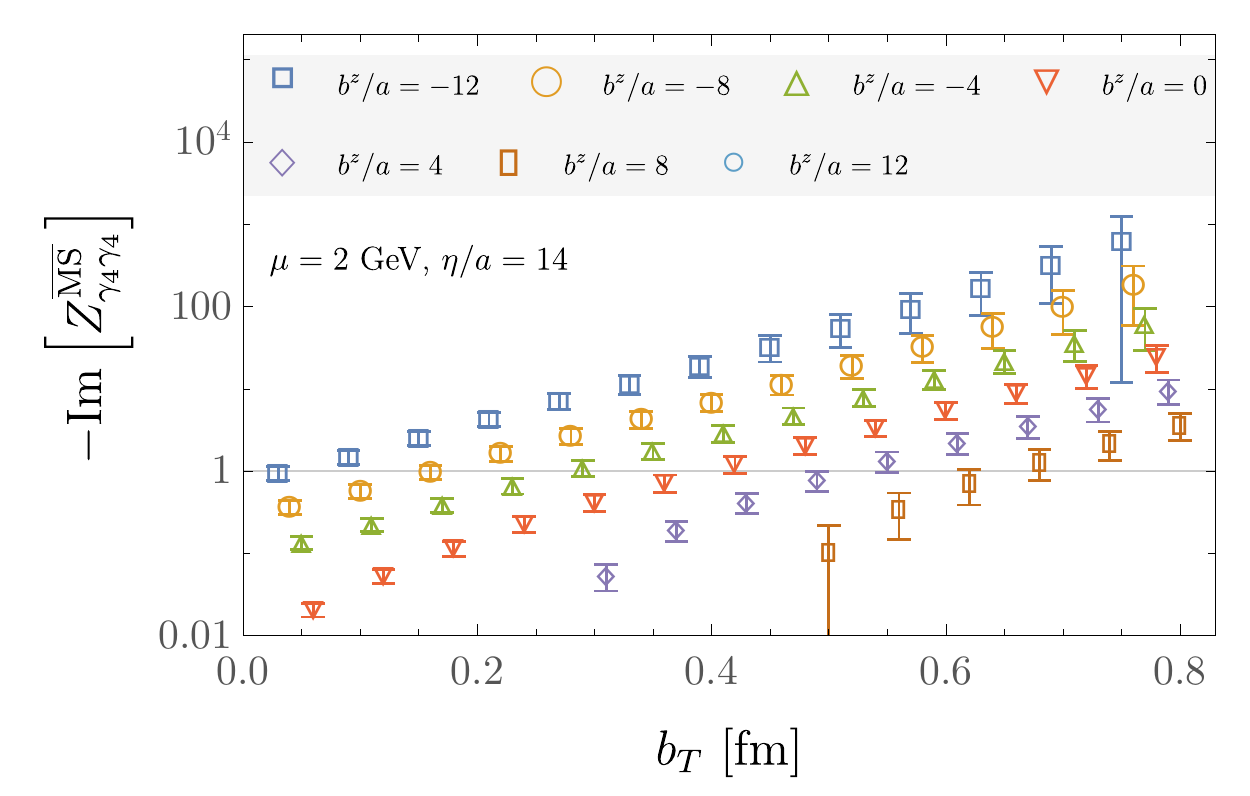}
        \label{fig:ZIvsBT} 
        }
        \caption{\label{fig:ZMSg4g4}
           Diagonal $\overline{\text{MS}}$ renormalization constants $Z^{\overline{\text{MS}}}_{\mathcal{O}_{\gamma_4\gamma_4}}(\mu = 2~\text{GeV})$, for staple-shaped operators with for $\eta/a=14$ and different geometries, computed on the $E_{32}$ ensemble.
        Points are shown with a small relative offset horizontally for clarity. Note that statistical noise grows with the length of the staple: $\eta + b_T + |\eta - b^z|$.}
\end{figure*}

\begin{figure*}
    \subfigure[]{
        \centering
        \includegraphics[width=0.46\textwidth]{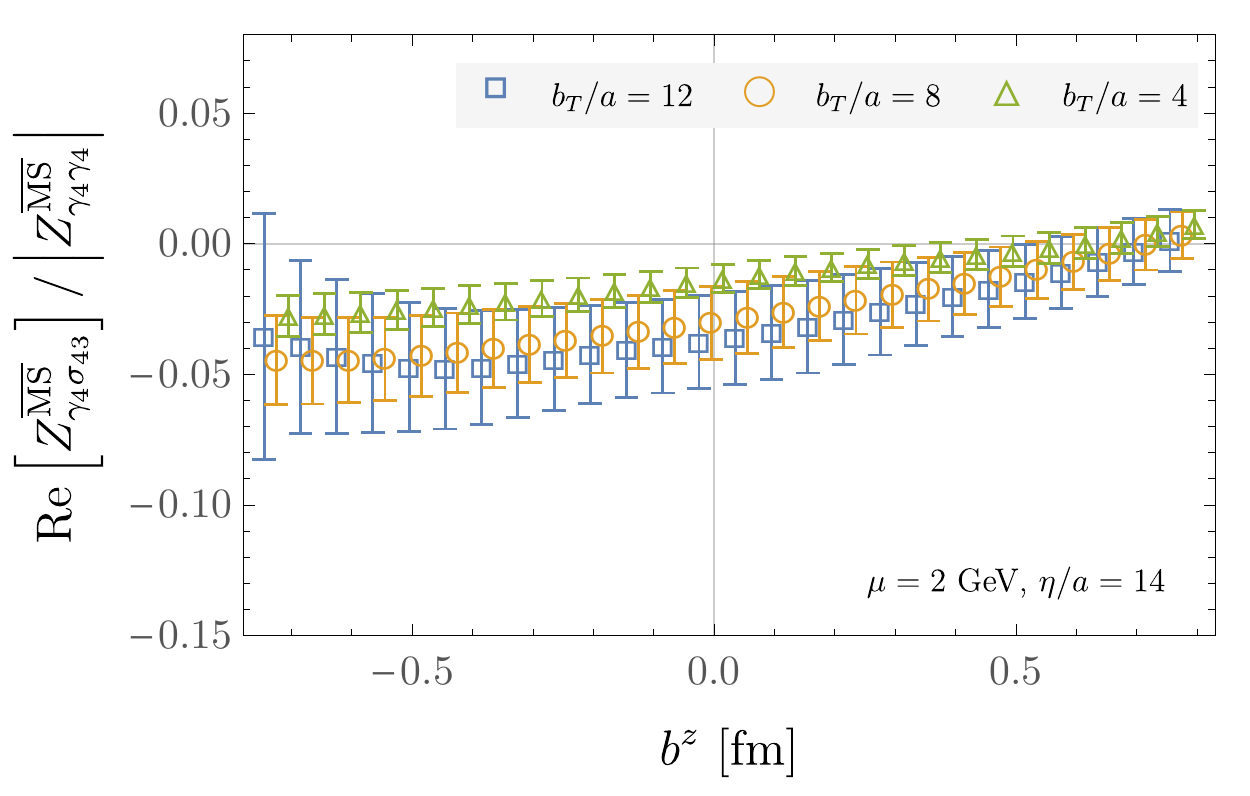}
        \label{fig:Z32RvsBzPlotMixPT}
        }\quad
    \subfigure[]{
        \centering
        \includegraphics[width=0.46\textwidth]{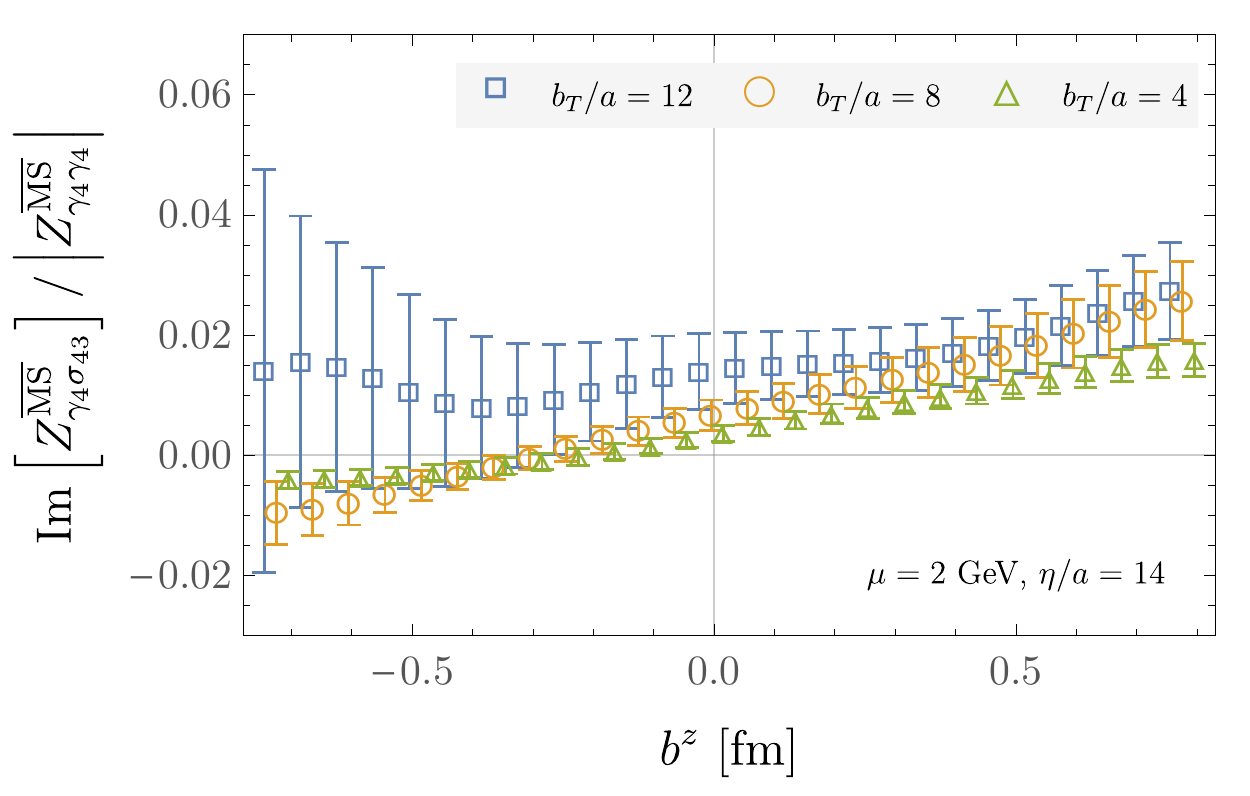}
        \label{fig:Z32RvsBzPlotMixPT}
        }\quad
    \subfigure[]{
        \centering
        \includegraphics[width=0.46\textwidth]{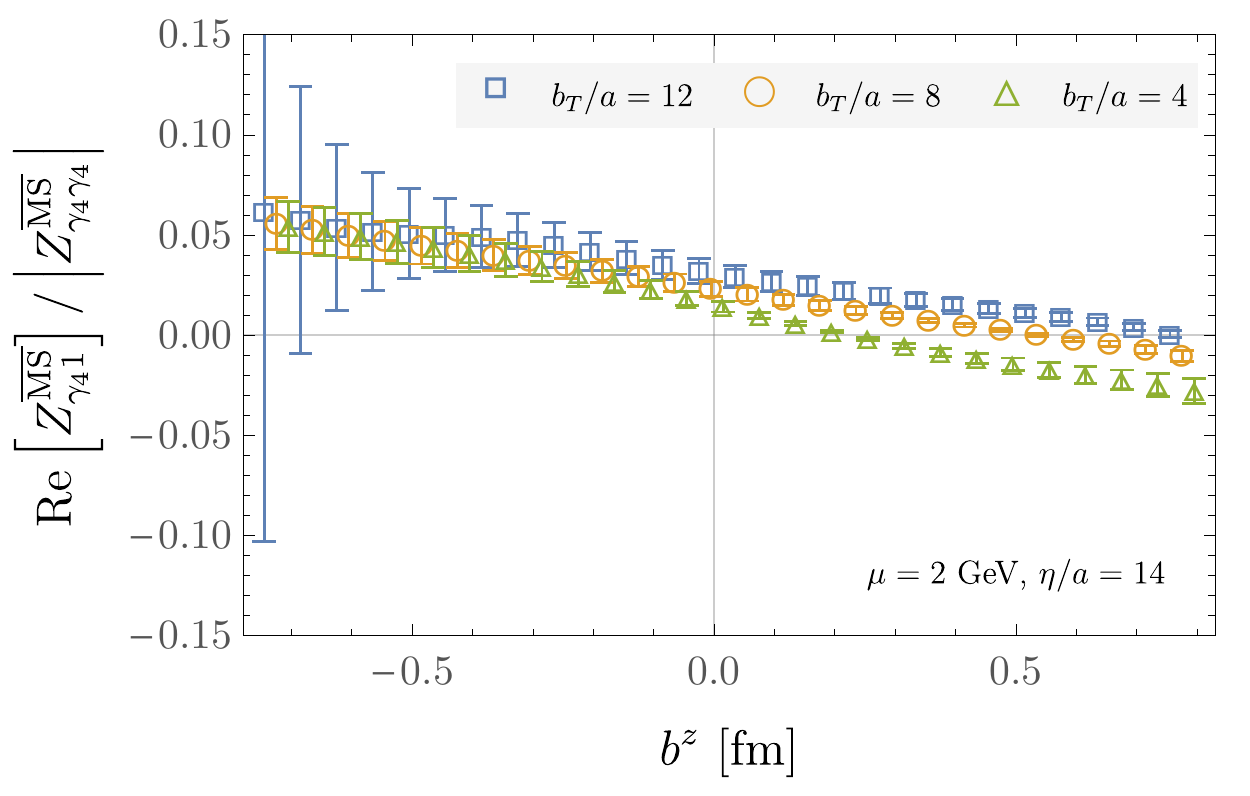}
        \label{fig:Z32RvsBzPlotMix}
        }\quad
    \subfigure[]{
        \centering
        \includegraphics[width=0.46\textwidth]{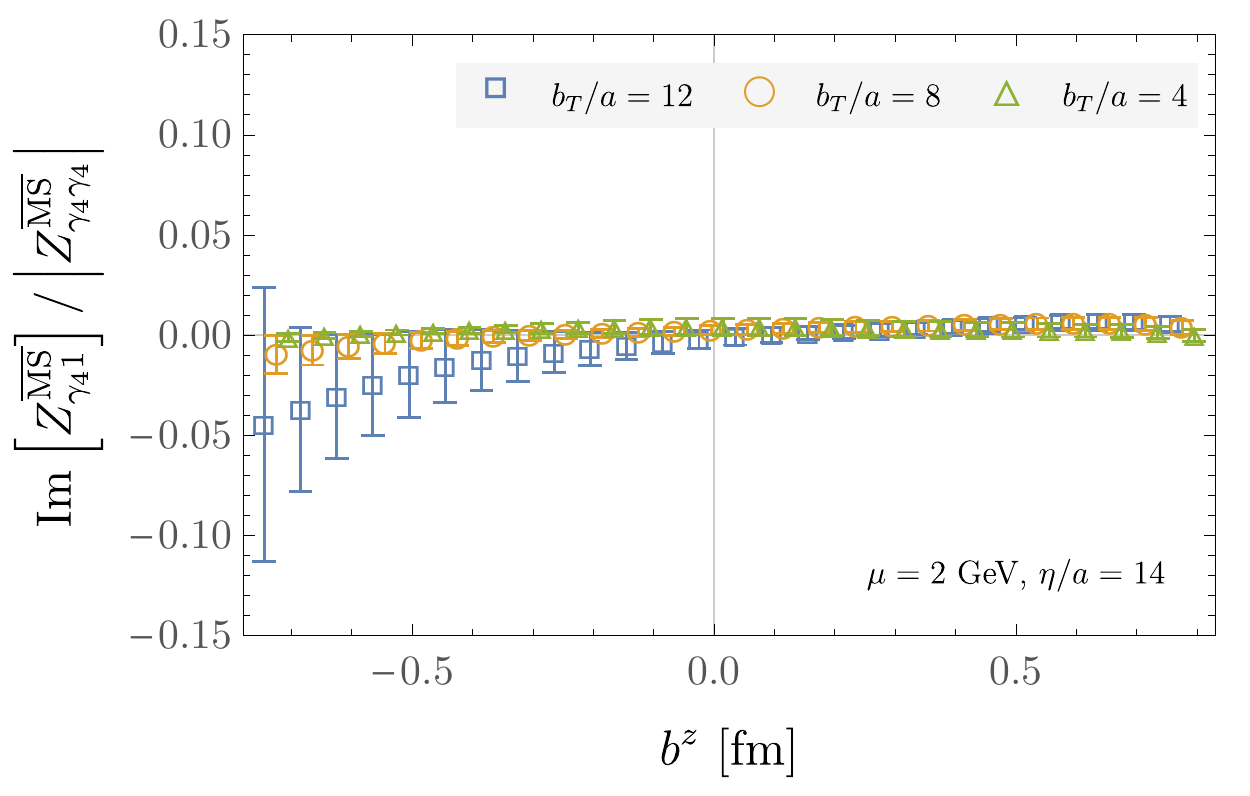}
        \label{fig:Z32IvsBzPlotMix}
        }
        \caption{\label{fig:ZMSod}
        Ratios of off-diagonal and diagonal $\overline{\text{MS}}$ renormalization constants describing operator mixing, for quark bilinear operators with staple-shaped Wilson lines with $\eta/a=14$, $\mu = 2$ GeV, and different staple geometries, computed on the $E_{32}$ ensemble. Points are shown with a small relative offset horizontally for clarity.}
\end{figure*}

\begin{figure*}
    \subfigure[]{
        \centering
        \includegraphics[width=0.46\textwidth]{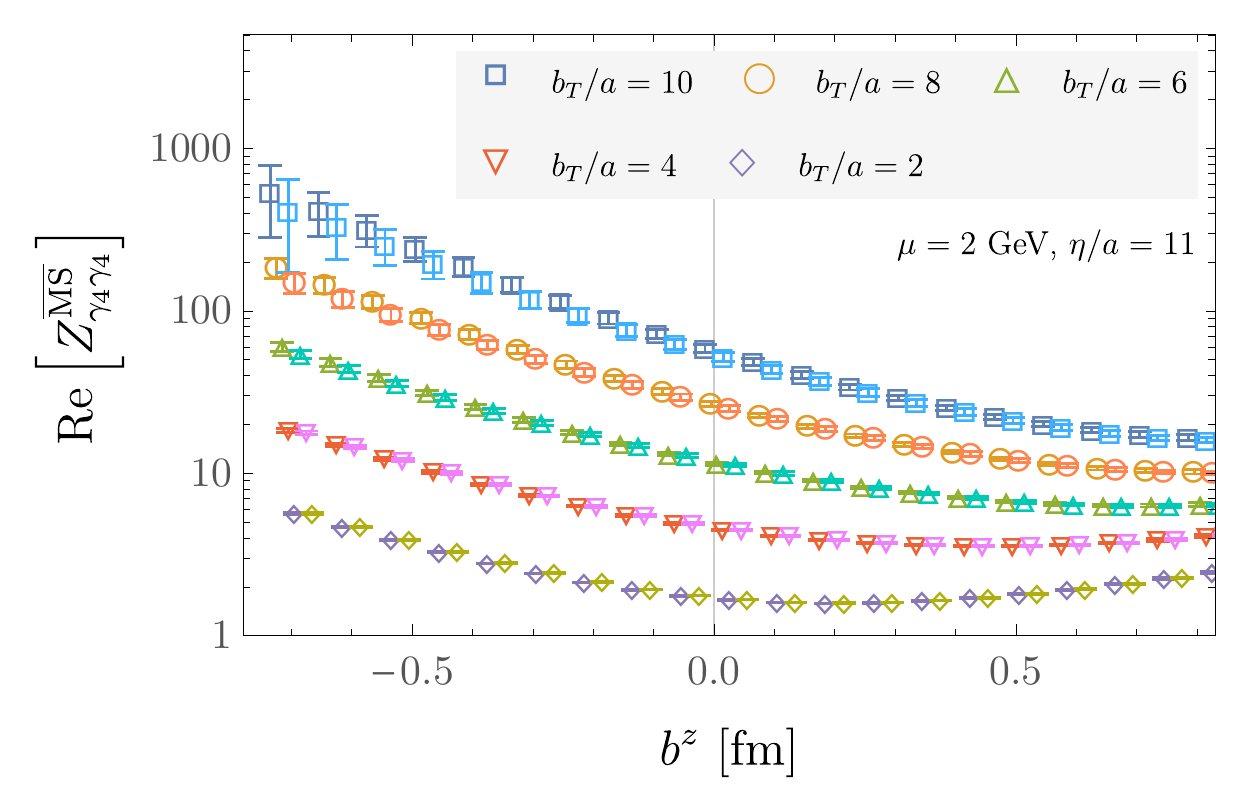}
        }\quad
    \subfigure[]{
        \centering
        \includegraphics[width=0.46\textwidth]{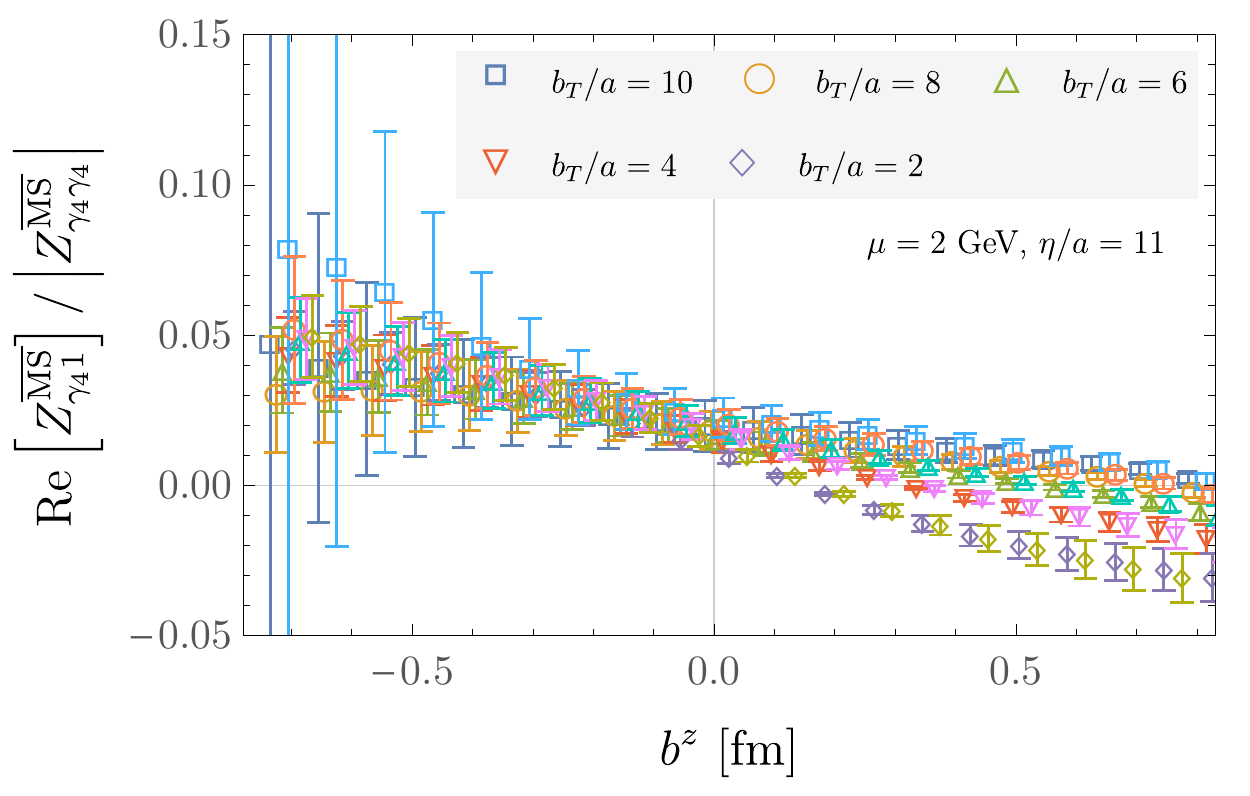}
        }\quad
        \caption{\label{fig:Zlight}  $\overline{\text{MS}}$ renormalization constants $Z^{\overline{\text{MS}}}_{\mathcal{O}_{\gamma_4\gamma_4}}(\mu = 2~\text{GeV})$ for quark bilinear operators with staple-shaped Wilson lines with $\eta / a = 11$ and different staple geometries for the $E_{24}$ ensemble and for a lighter quark mass corresponding to $m_\pi \sim 340$ MeV.  The left figure shows diagonal elements of the renormalization matrices, while the right figure shows ratios of off-diagonal to diagonal elements.
       Points are shown with a small relative offset horizontally for clarity. }
\end{figure*}

\begin{figure*}[]
    \subfigure[]{
        \centering
        \includegraphics[width=0.46\textwidth]{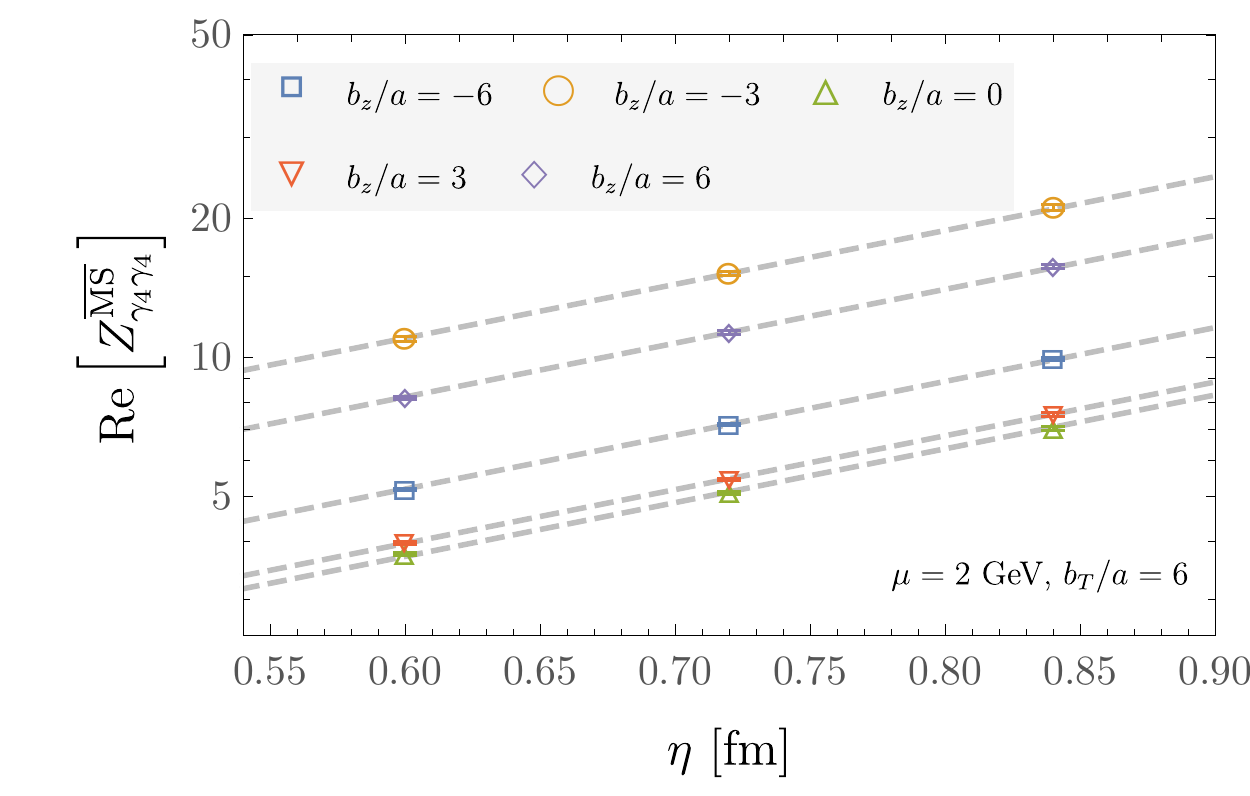}
        }\quad
    \subfigure[]{
        \centering
        \includegraphics[width=0.46\textwidth]{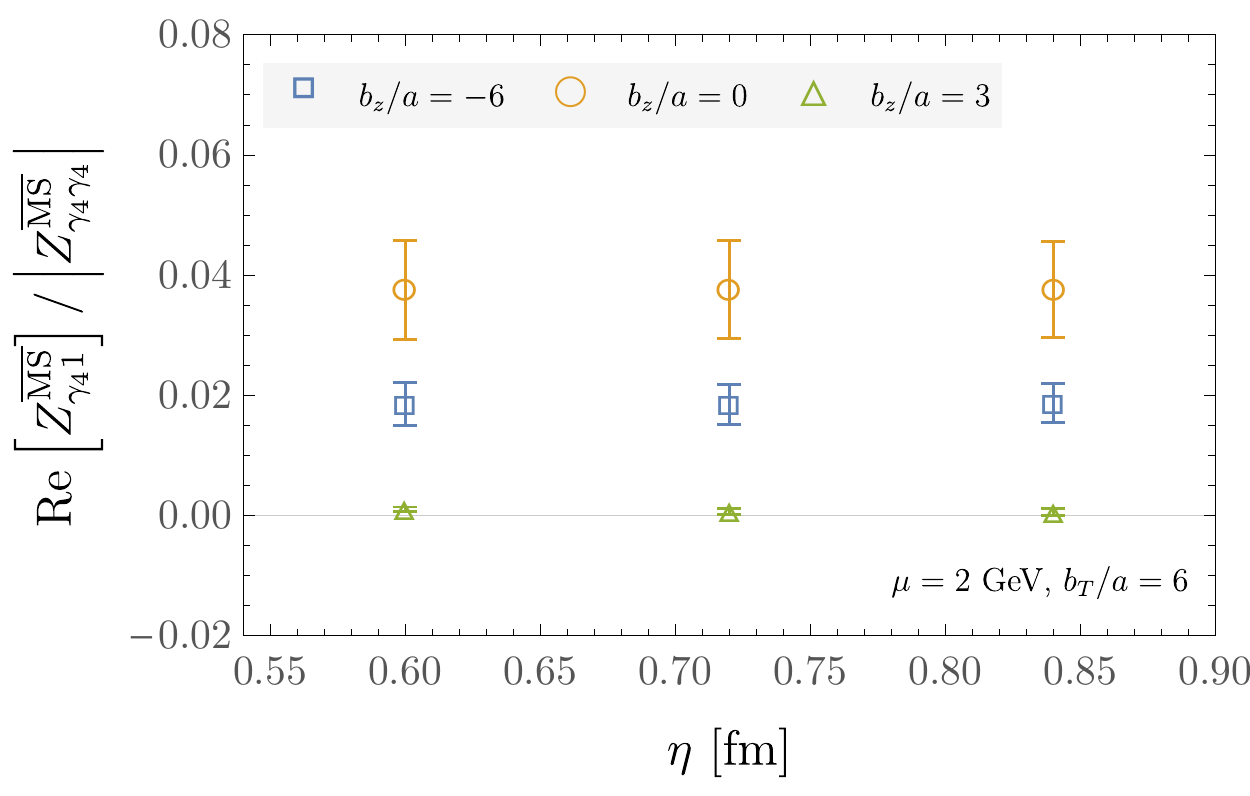}
        }\quad
        \caption{\label{fig:Zvseta} $\eta$ dependence of the $\overline{\text{MS}}$ renormalization constants $Z^{\overline{\text{MS}}}_{\mathcal{O}_{\Gamma\Gamma'}}(\mu = 2~\text{GeV})$ for the $E_{32}$ ensemble with $b_T / a = 6$, $\mu = 2$ GeV, and different $b^z$ as indicated. The left figure shows diagonal elements of the renormalization matrices, while the right figure shows ratios of off-diagonal to diagonal elements. Dashed lines show fits to the exponential dependence on $\eta$ in Eq.~\eqref{eq:explatt}, with independent normalization for each $\eta$ and a single common exponent $\delta =  0.08051(71)$.}
\end{figure*}

\begin{figure*}[]
    \subfigure[]{
        \centering
        \includegraphics[width=0.46\textwidth]{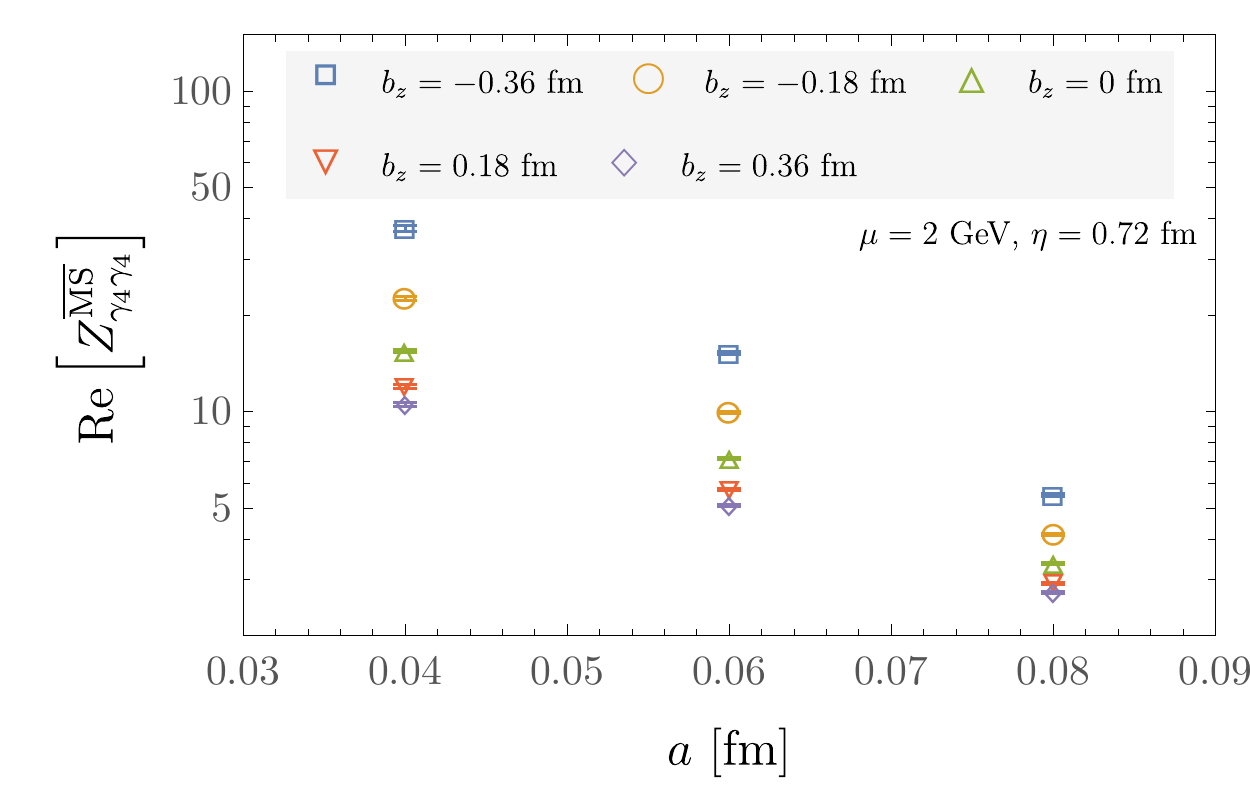}
        }\quad
    \subfigure[]{
        \centering
        \includegraphics[width=0.46\textwidth]{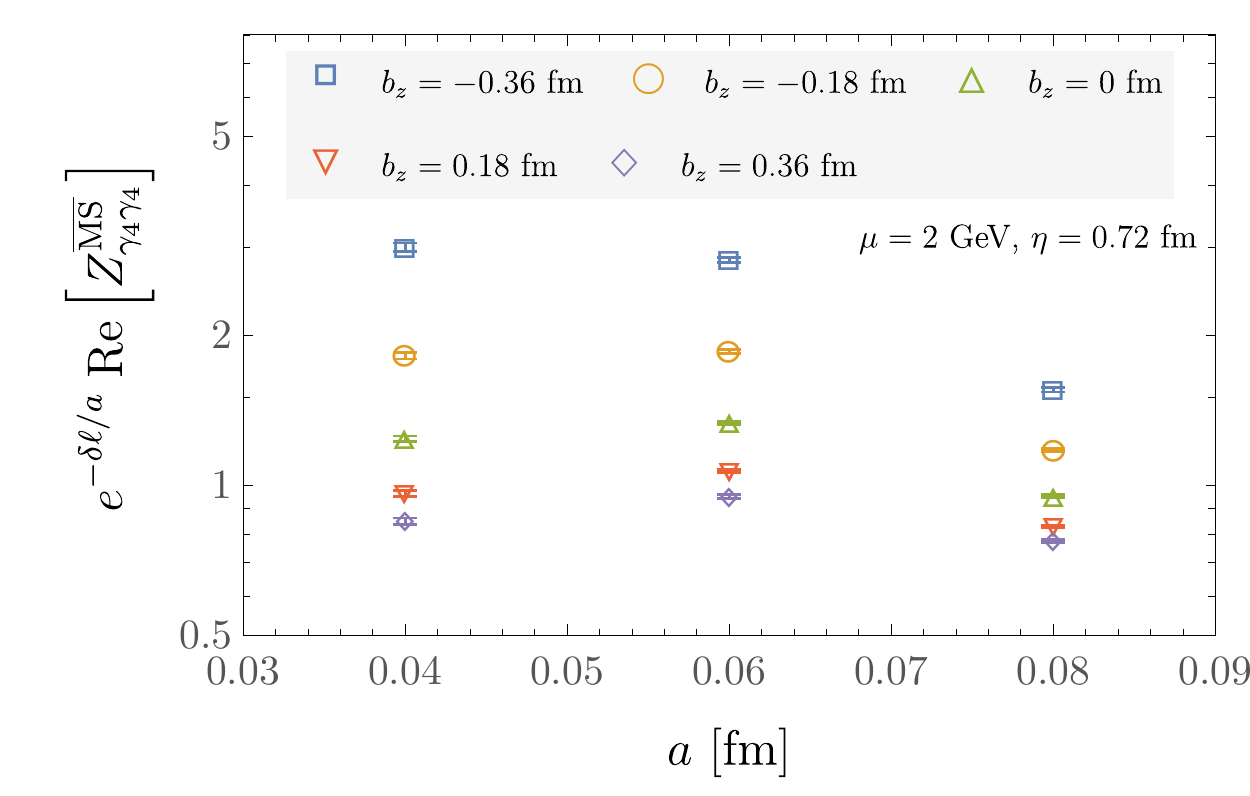}
        }\quad
        \caption{\label{fig:Zvsa} Lattice spacing dependence of $\overline{\text{MS}}$ renormalization constants $Z^{\overline{\text{MS}}}_{\mathcal{O}_{\gamma_4\gamma_4}}(\mu = 2~\text{GeV})$, for quark bilinear operators with Wilson line geometry defined by $\eta = 0.72$ fm, $b_T = 0.36$ fm, $\mu = 2$ GeV, and different $b^z$ as indicated. The left figure shows results for $Z^{\overline{\text{MS}}}_{\mathcal{O}_{\gamma_4\gamma_4}}(\mu = 2~\text{GeV})$, while the right figure shows the same results rescaled by $e^{-\delta \ell /a}$, with the best-fit value of $\delta = 0.10$ taken from a simultaneous fit to all three ensembles by Eq.~\eqref{eq:explatt}, as described in the text.}
\end{figure*}

\begin{figure*}[]
    \subfigure[]{
        \centering
        \includegraphics[width=0.46\textwidth]{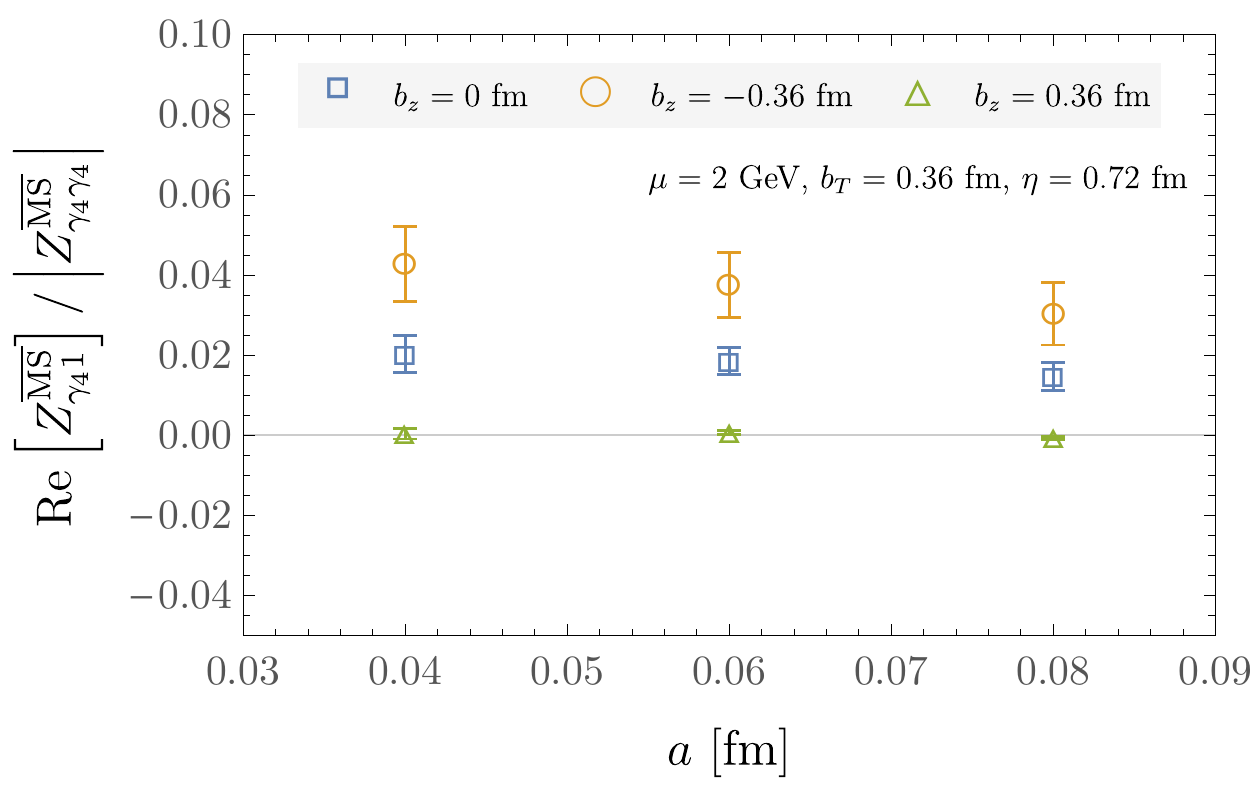}
        }\quad
    \subfigure[]{
        \centering
        \includegraphics[width=0.46\textwidth]{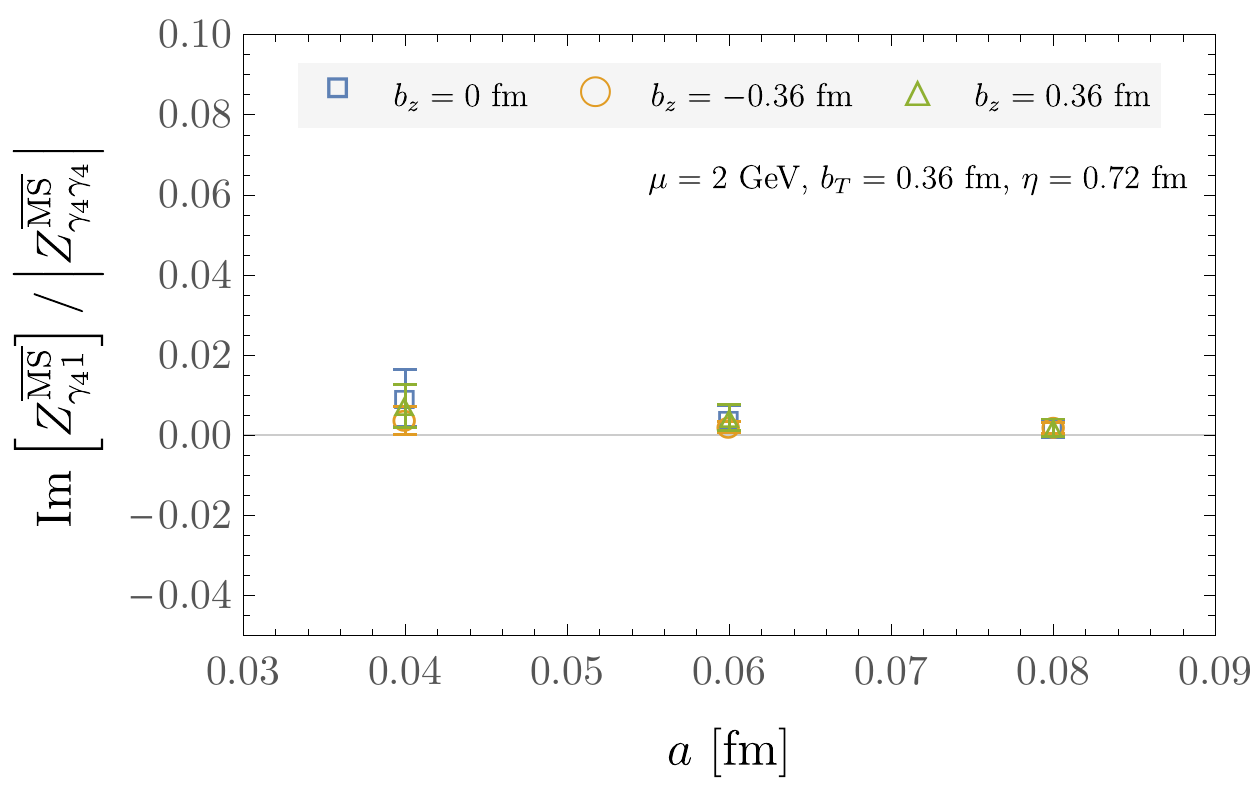}
        }\quad
        \caption{Ratios of off-diagonal and diagonal $\overline{\text{MS}}$ renormalization constants for operators with $\eta = 0.72$ fm, $b_T = 0.36$ fm, $\mu = 2$ GeV, and different $b^z$, as a function of lattice spacing $a$. \label{fig:ZvsaMix}}
\end{figure*}

Rather than performing uncorrelated fits to correlated results, weighted averages are used to remove residual $p_R$ dependence,
\begin{equation}
   \begin{split}
      Z^{\overline{\text{MS}}}_{\mathcal{O}_{\gamma_4\Gamma}}(\mu) &= \sum_n w_n Z^{\overline{\text{MS}}}_{\mathcal{O}_{\gamma_4\Gamma}}(\mu,p_R^n)\,, \\
      \delta_{\text{stat}} Z^{\overline{\text{MS}}}_{\mathcal{O}_{\gamma_4\Gamma}}(\mu)^2 &= \sum_n w_n \delta Z^{\overline{\text{MS}}}_{\mathcal{O}_{\gamma_4\Gamma}}(\mu,p_R^n)^2\,, \\
      \delta_{\text{sys}} Z^{\overline{\text{MS}}}_{\mathcal{O}_{\gamma_4\Gamma}}(\mu)^2 &= \sum_n \left( Z^{\overline{\text{MS}}}_{\mathcal{O}_{\gamma_4\Gamma}}(\mu) - Z^{\overline{\text{MS}}}_{\mathcal{O}_{\gamma_4\Gamma}}(\mu,p_R^n) \right)^2\,, \\
      \delta Z^{\overline{\text{MS}}}_{\mathcal{O}_{\gamma_4\Gamma}}(\mu)^2 &= \delta_{\text{stat}} Z^{\overline{\text{MS}}}_{\mathcal{O}_{\gamma_4\Gamma}}(\mu)^2 + \delta_{\text{sys}} Z^{\overline{\text{MS}}}_{\mathcal{O}_{\gamma_4\Gamma}}(\mu)^2,
   \end{split}\label{eq:weightedave}
\end{equation}
where the weights are chosen to sum to unity and to be proportional to the inverse variance of the result for each momentum:
\begin{equation}
      w_n = \frac{\tilde{w}_n}{\sum_n \tilde{w}_n}, \hspace{5mm}
      \tilde{w}_n = \frac{1}{\delta Z^{\overline{\text{MS}}}_{\mathcal{O}_{\gamma_4\Gamma}}(\mu,p_R^n)^2}.
\label{eq:weights}
\end{equation}
The central value of this weighted average is identical to the central value of an uncorrelated fit and ensures that the fit is constrained most heavily by the most precise data.
The inverse variance of this weighted average is the average inverse variance of the data, while the inverse variance of an uncorrelated $\chi^2$-minimization fit is equal to the same quantity times the number of data points.
Uncorrelated fits to correlated data therefore lead to a spurious reduction in the uncertainty of the fit result that is avoided by Eq.~\eqref{eq:weightedave}.
The systematic uncertainty term in Eq.~\eqref{eq:weightedave} is included to reflect the uncertainty arising from unresolved discretization and nonperturbative effects. 
The resulting systematic error on $Z^{\overline{\text{MS}}}_{\mathcal{O}_{\gamma_4\gamma_4}}$ is  $< 15\%$ in all cases; for all but the largest Wilson line extents the systematic uncertainty is $\lesssim 2\%$.
Similar results hold for $Z^{\overline{\text{MS}}}_{\mathcal{O}_{\gamma_4\Gamma}}$ with $\Gamma \neq \gamma_4$ apart from cases where $Z^{\overline{\text{MS}}}_{\mathcal{O}_{\gamma_4\Gamma}}$ is consistent with zero.

Figure~\ref{fig:matching} shows a representative example of this weighted averaging procedure for an asymmetric staple operator with $\eta/a = 10$, $b_T/a = 3$, and $b^z/a = 4$, computed on ensemble $E_{32}$.
For this example and in general, the $\overline{\text{MS}}$ renormalization constant $Z^{\overline{\text{MS}}}_{\mathcal{O}_{\gamma_4\Gamma}}(\mu, p_R)$ is more consistent with a constant and has smaller systematic uncertainties in a weighted average than $Z^{\text{$\RI$}}_{\mathcal{O}_{\gamma_4\Gamma}}(p_R)$, which indicates that one-loop matching accounts for some of the $p_R$-dependence of the bare vertex function. 
Results for operators with displacements in the $x-z$ and $y-z$ planes, where $x$ and $y$ are the  directions transverse to the staple extent $\eta$, are fit independently and found to be consistent within uncertainties, and the renormalization constants for operators of different shapes
are found to be relatively smooth functions of the staple geometry parametrized by $b^z$, $b_T$, and $\eta$. 
Samples of these results are shown for the diagonal renormalization constant $Z^{\overline{\text{MS}}}_{\mathcal{O}_{\gamma_4\gamma_4}}(\mu)$ in Figure~\ref{fig:ZMSg4g4}. 
Here and throughout, $\mu = 2~\text{GeV}$ is used as a reference scale.
The off-diagonal terms $Z^{\overline{\text{MS}}}_{\mathcal{O}_{\gamma_4\Gamma}}(\mu)$ with $\Gamma \neq \gamma_4$ describing operator mixing indicate that such mixing is a percent-level effect for operators with small Wilson lines, but grows to become a $5-10\%$ effect for the largest Wilson lines studied.
A representative set of off-diagonal mixing results are shown in Figure~\ref{fig:ZMSod}.

In order to study the quark mass dependence of $Z^{\overline{\text{MS}}}_{\mathcal{O}_{\gamma_4\gamma_4}}$, calculations on the $E_{24}$ ensemble are repeated using a second quark mass corresponding to $m_\pi \sim 340$ MeV.
For all $Z^{\overline{\text{MS}}}_{\mathcal{O}_{\gamma_4\Gamma}}$, results for $m_\pi \sim 340$ MeV are found to be consistent within uncertainties with those calculated using $m_\pi \sim 1.2$ GeV, as shown in Figure~\ref{fig:Zlight}.
This suggests that the large quark mass used in this work does not significantly affect results for $Z^{\overline{\text{MS}}}_{\mathcal{O}_{\gamma_4\Gamma}}$.
Before averaging over momentum, statistically significant differences between $m_\pi \sim 1.2$ GeV and $m_\pi \sim 340$ MeV results can be seen at the smallest momenta considered here, which is consistent with expectations that renormalization factors include nonperturbative quark mass effects proportional to $m_q \left< \overline{q} q\right> / p^4$ that vanish at large momentum~\cite{Boucaud:2000nd,Boucaud:2001st,Boucaud:2005rm,Aoki:2007xm,Constantinou:2010gr,Gockeler:2010yr,Blossier:2010vt,Blossier:2014kta}.
After averaging over momentum, results with $m_\pi \sim 1.2$ GeV and $m_\pi \sim 340$ MeV are consistent within combined statistical and systematic uncertainties. 

For operators constructed from long Wilson lines, the renormalization factors are found to depend approximately exponentially on the Wilson line extent. 
In particular, $Z^{\overline{\text{MS}}}_{\mathcal{O}_{\gamma_4\gamma_4}}$ with fixed $b_T$ and $b^z$ has an approximately exponential dependence on $\eta$ as shown in Figure~\ref{fig:Zvseta}. 
This $\eta$-dependence should cancel corresponding $\eta$-dependence in bare matrix elements, resulting in approximately $\eta$-independent $\MS$ renormalized matrix elements.
Similar exponential dependence on the Wilson line extent is seen in the $b^z$ and $b_T$ dependence of $Z^{\overline{\text{MS}}}_{\mathcal{O}_{\gamma_4\gamma_4}}$ in Figure~\ref{fig:ZMSg4g4} for large $b_T$ and large $-b^z$.
At smaller values of $b_T$ and $-b^z$, additional structure beyond simple exponential dependence on the Wilson line length is demonstrated by the curvature visible in Figure~\ref{fig:ZMSg4g4}.

This behaviour is consistent with expectations from perturbation theory: nonlocal operators with Wilson lines include $1/a$ divergences arising at one-loop in lattice perturbation theory, which can be resummed yielding exponential dependence on $1/a$~\cite{Dotsenko:1979wb,Craigie:1980qs,Dorn:1986dt,Xiong:2013bka,Ji:2015jwa,Chen:2016fxx,Ishikawa:2016znu}. For quark bilinears with symmetric staple-shaped Wilson lines, these $1/a$ divergences were explicitly calculated in Ref.~\cite{Constantinou:2019vyb}.
The $a\rightarrow 0$ divergence of $Z^{\overline{\text{MS}}}_{\mathcal{O}_{\gamma_4\gamma_4}}$ predicted by lattice perturbation theory can be parametrized as
\begin{equation}
   \begin{split}
      Z^{\overline{\text{MS}}}_{\mathcal{O}_{\gamma_4\Gamma}} = \mathcal{A} \; e^{ \delta \, \ell/a}\left(1 + \ldots \right),
   \end{split}\label{eq:explatt}
\end{equation}
where $\ell = \eta + b_T + |\eta - b^z|$ is the length of the Wilson line, $\mathcal{A}$ is a constant, and omitted terms include logarithmically divergent contributions as $a\rightarrow 0$ as well as $O(a)$ terms that vanish in the continuum limit.
The coefficient $\delta$ of the $1/a$ one-loop divergence depends on the lattice action and in particular on the smearing prescription applied to the gauge field; for the flowed gauge field ensembles used here, $\delta$ is treated as a free parameter that can be fit to nonperturbative results.

The $\eta$ dependence of $Z^{\overline{\text{MS}}}_{\mathcal{O}_{\gamma_4\gamma_4}}$ is described accurately by Eq.~\eqref{eq:explatt} for fixed $b_T$, $b^z$, and $a$.
Treating $\mathcal{A}$ as a $b^z$-dependent normalization factor, uncorrelated $\chi^2$-minimization fits to the $E_{32}$ ensemble results with $\eta/a=\{10,12,14\}$, $b_T/a=6$, and $b^z/a = \{-6,-3,0,3,6\}$, shown in Figure~\ref{fig:Zvseta}, yield $\delta = 0.08051(71)$ with uncertainties estimated using bootstrap resampling, and $\chi^2/\text{dof} = 0.53$ with 9 degrees of freedom. The $\eta$-dependence of results computed on the $E_{24}$ and $E_{48}$ ensembles can be fit using Eq.~\eqref{eq:explatt} in a similar way; however, fitting the $\eta$-dependence of results for all three ensembles simultaneously results in a $\chi^2/\text{dof}$ of over 500. This indicates that there is significant remaining $a$-dependence that is not captured by this functional form. Nevertheless, taking $\delta$ from this combined fit to all three ensembles and rescaling by taking products with $e^{-\delta \ell/a}$ largely removes the power-law divergences in the renormalization factors, as shown in Figure~\ref{fig:Zvsa}.

Ratios of off-diagonal and diagonal elements of the renormalization matrices $Z^{\overline{\text{MS}}}_{\mathcal{O}_{\gamma_4\Gamma}} / \left| Z^{\overline{\text{MS}}}_{\mathcal{O}_{\gamma_4\gamma_4}} \right|$ are seen to have mild $a$-dependence, as shown in Figure~\ref{fig:ZvsaMix}.
This is consistent with general arguments that the $a\rightarrow 0$ divergence structure of $Z^{\overline{\text{MS}}}_{\mathcal{O}_{\gamma_4\Gamma}}$ does not depend on $\Gamma$ discussed in Ref.~\cite{Musch:2010ka}.

\section{Summary}

In this work, the nonperturbative $\RI$ renormalization of staple-shaped Wilson line operators, as relevant to lattice QCD studies of transverse-momentum-dependent parton distribution functions, is investigated for the first time. 
The renormalization factors are computed nonperturbatively for a basis of nonlocal quark bilinear operators with a wide range of transverse and longitudinal separations in quenched QCD with three different lattice spacings, namely 0.04, 0.06, and 0.08~fm, and a single lattice volume, $L\sim 2$~fm.
Renormalization factors are found to depend exponentially on the length of the Wilson line in lattice units, as expected from perturbation theory, although additional dependence on the shape of the Wilson line is clearly visible.
Quark mass dependence is found to be negligible compared to uncertainties from statistical noise and lattice artifacts, for quark masses corresponding to $m_\pi=\{0.4,1.2\}$~GeV.

Mixing between nonlocal quark bilinears with different Dirac operator structures is observed to be larger than the corresponding mixing between local quark bilinears; this mixing can not be neglected in studies of nonlocal quark bilinears targeting precision better than the $10\%$ level.
These operator mixing effects show mild lattice spacing dependence, with the effects typically found to be larger both for finer discretization scales and for operators built from Wilson lines with longer staple extents.
While the mixing patterns predicted by one-loop lattice perturbation theory are observed, additional mixing effects that are as, or even more, significant than the predicted mixings, are also present for nonlocal quark bilinear operators with both straight and staple-shaped Wilson lines. For operators with straight Wilson lines relevant to calculations of quasi PDFs, block-diagonal structure is observed in the mixing patterns, while for operators with staple-shaped Wilson lines a dense mixing pattern is observed.  
This result also suggests that caution should be used in the application of one-loop lattice perturbation theory to remove momentum-dependent discretization effects associated with RI/MOM type schemes, for calculations of quasi beam functions in a framework similar to that studied here.
The results of this work allow bare matrix elements for a basis of non-local quark bilinear operators with staple-shaped Wilson lines to be renormalized, with the mixing between operators with different Dirac structures fully accounted for.
This completes a critical step towards the systematic extraction of TMDPDFs, and also TMD distribution amplitudes, from lattice QCD.

\section*{Acknowledgements}

The authors thank Will Detmold, Markus Ebert, Jeremy Green, Andrew Pochinsky, and Iain Stewart for helpful discussions, and Michael Endres for providing the gauge field configurations used in this project.
Calculations were performed using the Qlua~\cite{qlua} and Chroma~\cite{Edwards:2004sx} software libraries.
This work is supported in part by the U.S.~Department of Energy, Office of Science, Office of Nuclear Physics, under grant Contract Number DE-SC0011090, DE-SC0012704 and within the framework of the TMD Topical Collaboration. PES is additionally supported by the National Science Foundation under CAREER Award 1841699. MLW is additionally supported by an MIT Pappalardo fellowship.
Computations for this work used resources of the National Energy Research Scientific Computing Center (NERSC), a U.S. Department of Energy Office of Science User Facility operated under Contract No. DE-AC02-05CH11231, as well as the Extreme Science and Engineering Discovery Environment (XSEDE), which is supported by National Science Foundation grant number ACI-1548562, and facilities of the USQCD Collaboration, which are funded by the Office of Science of the U.S. Department of Energy.

\appendix

\begin{figure}[!t]
    \centering
    \includegraphics[width=\columnwidth]{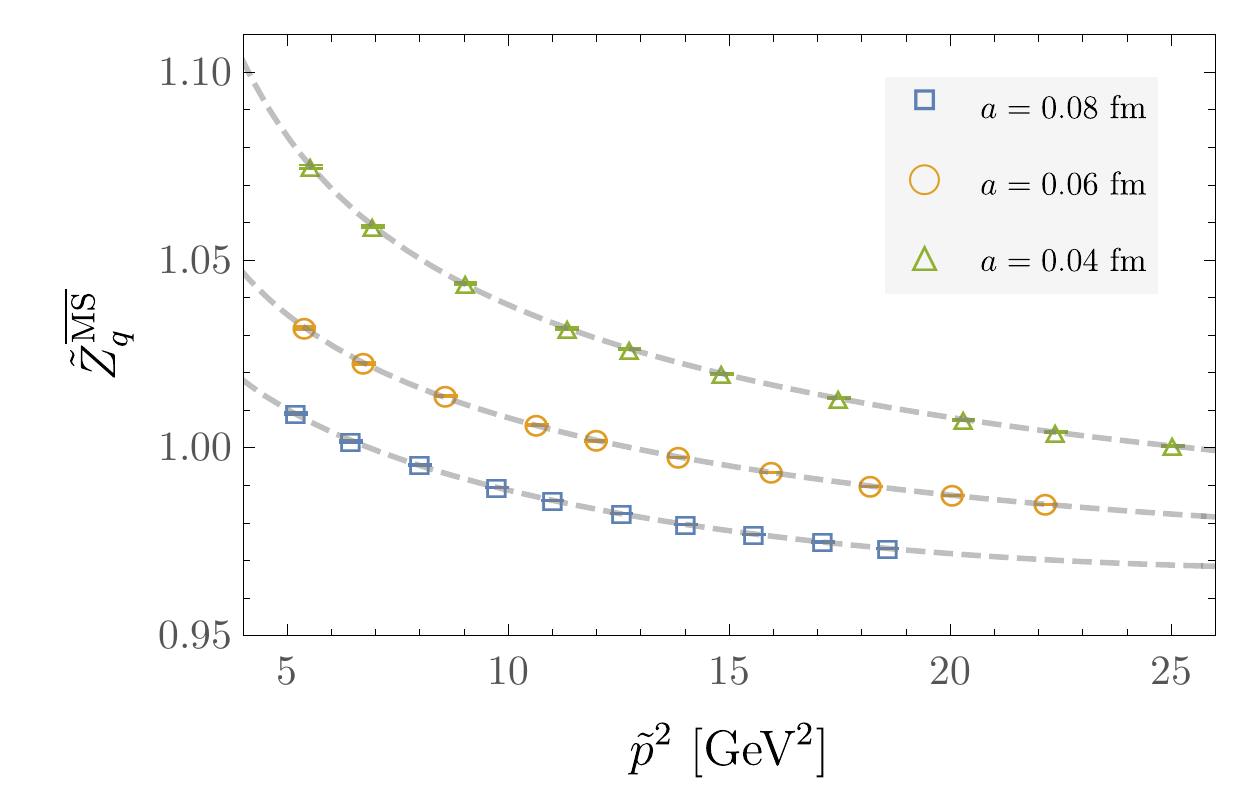}
    \caption{Numerical results for the subtracted quark wavefunction renormalization constant $\tilde{Z}_q^{\MS}$ defined in Eq.~\eqref{eq:Zqscalingtilde} from all three ensembles. Bands show the results of independent fits to Eq.~\eqref{eq:Zqscalingtilde} for each ensemble. 
    \label{fig:Zq} }
\end{figure}

\begin{figure*}
    \centering
    \includegraphics[width=\textwidth]{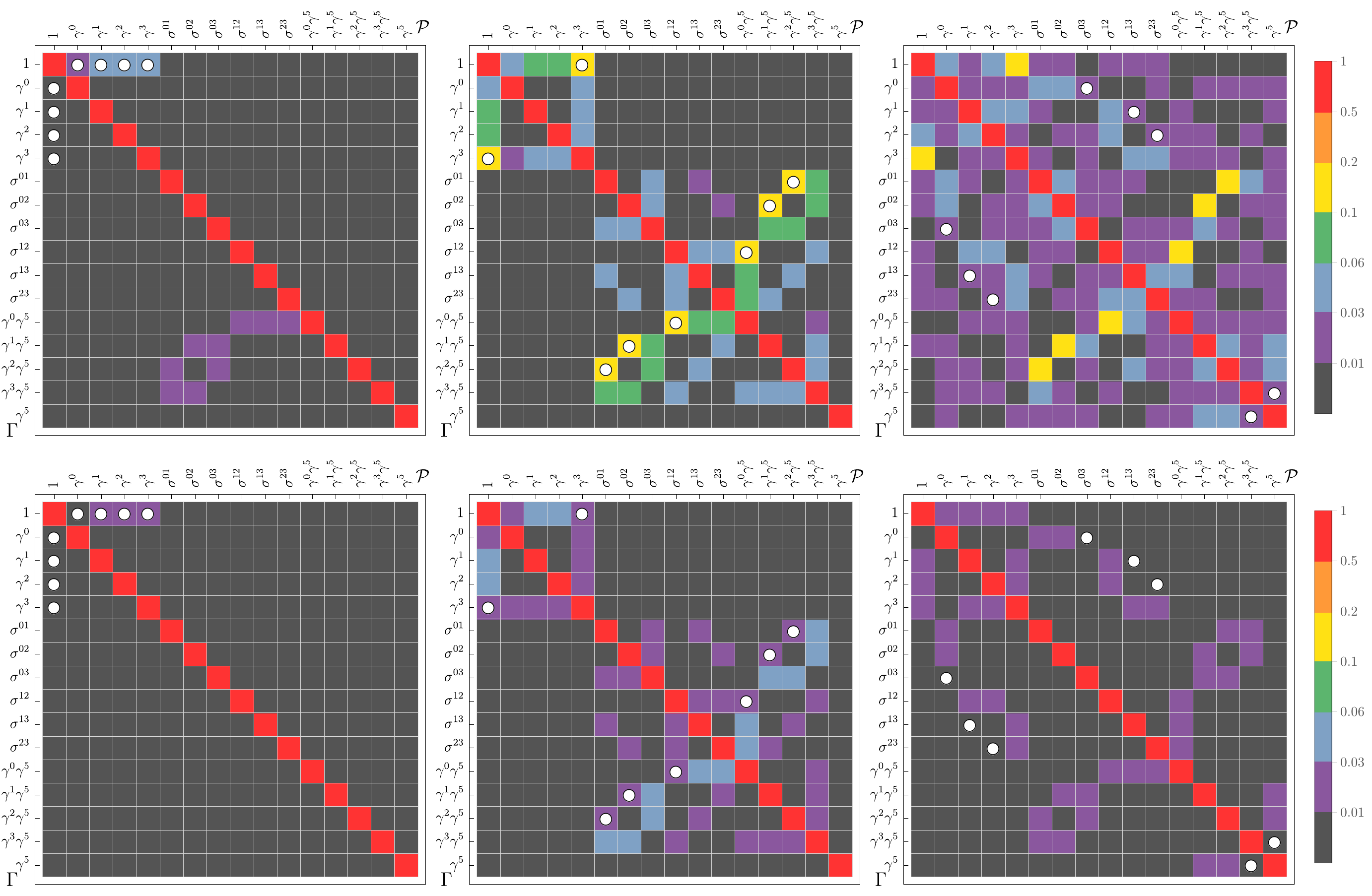}
    \caption{$\RI$ mixing pattern $\mathcal{M}^\text{$\RI$}_{\mathcal{O}_{\Gamma\mathcal{P}}}$ (Eq.~\eqref{eq:mixeq}) calculated for momentum $n^\mu = (4,4,4,4)$ on the $E_{24}$ ensemble with no Wilson flow (top row) and Wilson flow to $\mathfrak{t}=1$ in lattice units as in the main text (bottom row), applied to the gauge fields. Columns of figures from left to right show results for local operators, straight Wilson line operators ($b_T=0$) with extent $b^z/a=7$, and symmetric staple-shaped Wilson line operators ($b^z=0$) with extent $\eta/a=9$, $b_T/a=3$, respectively. The bare quark mass is tuned separated for flowed and unflowed gauge fields in order to give $m_\pi \sim 1.2$ GeV in both cases.
    White circles indicate the mixings predicted by one-loop lattice perturbation theory and symmetry arguments~\cite{Constantinou:2017sej,Chen:2017mie,Green:2017xeu,Constantinou:2019vyb}.}
    \label{fig:flow}
\end{figure*}

\section{Discretization effects}
\label{sec:disceffects}

To $O(a^2)$, and including the dominant $1/p^2$ effect, a model of discretization effects in $Z^{\overline{\text{MS}}}_{\mathcal{O}_{\gamma_4\Gamma}}$ for the momenta considered here can be expressed as~\cite{Blossier:2014kta}
\begin{equation}
   \begin{split}
      Z^{\overline{\text{MS}}}_{\mathcal{O}_{\gamma_4\Gamma}}(\mu,p) &= Z^{\overline{\text{MS}}}_{\mathcal{O}_{\gamma_4\Gamma}}(\mu) + c_1 \tilde{p}_z + c_2 \tilde{p}^2 + c_3 \tilde{p}_t^2 \\
                                                                   &\hspace{20pt} + c_4 \frac{\tilde{p}^{[4]}}{\tilde{p}^2} +  c_5 \tilde{p}^2 \ln(\tilde{p}^2) + \frac{d_1}{\tilde{p}^2} + \ldots ,
   \end{split}\label{eq:Zscaling}
\end{equation}
where $p^{[4]} = \sum_{\mu=1}^4 p_{\mu}^4$ and the parameters $c_i$ are $a$-dependent constants that can be extracted from fits to numerical data. On the right-hand side of this expression, momenta have been replaced with the momentum variable that arises in a discrete Fourier transform of the lattice action, namely
\begin{equation}\label{eq:ptil}
    \tilde{p}_\mu \equiv \frac{1}{a}\text{sin}(ap^\mu).
\end{equation}
Local operator renormalization factors and $Z_q$ have additional symmetry constraints leading to $c_1 = 0$ and $c_3 = 0$ up to negligible symmetry-breaking effects from the different extent of the lattice space and time directions.
This leads to the functional form
\begin{equation}
   \begin{split}
      Z^{\overline{\text{MS}}}_{q}(\mu,p) &= Z^{\overline{\text{MS}}}_{q}(\mu) + c_2 \tilde{p}^2 \\
                                                                   &\hspace{20pt} + c_4 \frac{\tilde{p}^{[4]}}{\tilde{p}^2} +  c_5 \tilde{p}^2 \ln(\tilde{p}^2) + \frac{d_1}{\tilde{p}^2} + \ldots .
   \end{split}\label{eq:Zqscaling}
\end{equation}
Results for $Z^{\overline{\text{MS}}}_{q}(\mu,p)$ are fit to Eq.~\eqref{eq:Zqscaling} for each ensemble.
Fit results for $c_4$ are used to remove rotationally non-invariant lattice artifacts as
\begin{equation}
   \begin{split}
      \tilde{Z}^{\overline{\text{MS}}}_{q}(\mu,p) &= Z^{\overline{\text{MS}}}_{q}(\mu,p) - c_4 \frac{\tilde{p}^{[4]}}{\tilde{p}^2}.
   \end{split}\label{eq:Zqscalingtilde}
\end{equation}
Figure~\ref{fig:Zq} shows numerical results for $\tilde{Z}^{\overline{\text{MS}}}_{q}(\mu = 2\text{ GeV},p)$ as well as the best fit to Eq.~\eqref{eq:Zqscaling} for each ensemble studied here.
Results for $c_2$, $c_4$, and $c_5$ are consistent within uncertainties for all three ensembles, as expected.
As discussed in Sec.~\ref{sec:numerics}, for the nonlocal operator renormalization constants $Z^{\overline{\text{MS}}}_{\mathcal{O}_{\gamma_4\Gamma}}(\mu,p)$, lattice artifacts cannot be clearly resolved, and simple constant fits are preferred over fits to Eq.~\eqref{eq:Zscaling} by information criteria.

\section{Wilson flow effects} \label{app:flow}

Wilson flow with a fixed flow-time $\mathfrak{t} = 1.0$ in lattice units is used in this work as a smearing prescription in order to improve signal-to-noise ratios of matrix elements including products of link operators.
To study the effect of Wilson flow on the results, calculations for a representative momentum, $n^\mu = (4,4,4,4)$ in lattice units, are repeated on the $E_{24}$ ensemble without Wilson flow applied to the gauge fields and with a value of $\kappa = 0.1403$ corresponding to $m_\pi = 1.22(2)$~GeV.
The resulting $\RI$ mixing patterns $\mathcal{M}^\text{$\RI$}_{\mathcal{O}_{\Gamma\mathcal{P}}}$, defined in Eq.~\eqref{eq:mixeq}, are shown in Figure~\ref{fig:flow} for this particular momentum with and without Wilson flow.
The off-diagonal elements of $\mathcal{M}^\text{$\RI$}_{\mathcal{O}_{\Gamma\mathcal{P}}}$ with Wilson flow are smaller than the results in Sec.~\ref{subsec:mixingpatterns} which show the maximum over 10 momentum from $E_{32}$.
In almost all cases, off-diagonal elements of $\mathcal{M}^\text{$\RI$}_{\mathcal{O}_{\Gamma\mathcal{P}}}$ without Wilson flow are larger than the corresponding mixings with flow.

For quark bilinear operators with straight Wilson lines, computed without Wilson flow, the mixings predicted by one-loop lattice perturbation theory are also the largest mixings nonperturbatively.
With Wilson flow, these mixings are reduced significantly and become smaller than mixings that are not predicted by one-loop lattice perturbation theory.
For symmetric staple-shaped Wilson line operators ($b^z = 0$) without Wilson flow, mixings between operators with Dirac structures $\Gamma$ and $\Gamma' \in \{\Gamma,\hat{z}\!\!\!/\}$ dominate over those predicted by one-loop lattice perturbation theory~\cite{Constantinou:2019vyb}.
With Wilson flow applied to the gauge fields, all mixings are significantly reduced. It will be interesting to see whether the flowed mixing patterns are postdicted by flowed one-loop lattice perturbation theory.

\bibliography{tmdreference}

\end{document}